\documentclass[useAMS,usenatbib]{mn2e}

\usepackage{epsfig}
\usepackage{amsmath}
\usepackage{amssymb}
\usepackage{natbib}

\usepackage{multicol}
\usepackage{multirow}

\usepackage[dvips]{color}

\def \mnras {MNRAS}
\def \apj {ApJ}

\def \apjl {ApJL}
\def \aap {A\&A}
\def \nat {Nature}
\def \araa {ARAA}

\def \pasj {PASJ}

\title[Low-level accretion in neutron-star X-ray binaries]{Low-level accretion in neutron-star X-ray binaries}

\author[Wijnands et al.]{
\parbox[t]{\textwidth}{
\raggedright
R. Wijnands$^{1}$\thanks{r.a.d.wijnands@uva.nl},  
N. Degenaar$^{2}$,
M. Armas Padilla$^{3,4}$\thanks{CEI Canarias: Campus Atl\'antico tricontinental},
D. Altamirano$^{5}\thanks{University Research Fellow}$,
Y. Cavecchi$^{1}$,
M. Linares$^{3,4}$,
A. Bahramian$^{6}$,
C. O. Heinke$^{6,7}$\thanks{Alexander von Humboldt Fellow}
}
\vspace{6pt}\\
$^{1}$Astronomical Institute Anton Pannekoek, 
University of Amsterdam, 
Postbus 94249, 1090 GE Amsterdam, The Netherlands\\
$^{2}$Institute of Astronomy, University of Cambridge, Madingley Road, Cambridge, CB3 OHA, UK\\
$^{3}$Instituto de Astrof{\'i}sica de Canarias, c/ V{\'i}a L{\'a}ctea s/n, E-38205 La Laguna, Tenerife, Spain\\
$^{4}$Departamento de Astrof\' \i sica, Universidad de la Laguna, La Laguna, E-38205, S/C de Tenerife, Spain\\
$^{5}$Physics \& Astronomy, University of Southampton, Southampton, Hampshire SO17 1BJ, UK\\
$^{6}$Department of Physics, University of Alberta, CCIS 4-183, Edmonton, AB T6G 2E1, Canada\\
$^{7}$Max-Planck-Institut f\"ur Radioastronomie, Auf dem H\"ugel 69, D-53121 Bonn, Germany\\
}

\begin{document}


\pagerange{\pageref{firstpage}--\pageref{lastpage}} \pubyear{0000}

\maketitle

\label{firstpage}

\begin{abstract} We search the literature for reports on the spectral
properties of neutron-star low-mass X-ray binaries when they have
accretion luminosities between $10^{34}$ and $10^{36}$ ergs s$^{-1}$,
corresponding to roughly 0.01\% - 1\% of the Eddington accretion rate
for a neutron star. We found that in this luminosity range the photon
index (obtained from fitting a simple absorbed power-law in the
0.5--10 keV range) increases with decreasing 0.5--10 keV X-ray
luminosity (i.e., the spectrum softens).  Such behaviour has been
reported before for individual sources, but here we demonstrate that
very likely most (if not all) neutron-star systems behave in a similar
manner and possibly even follow a universal relation. When comparing
the neutron-star systems with black-hole systems, it is clear that
most black-hole binaries have significantly harder spectra at
luminosities of $10^{34} -10^{35}$ erg s$^{-1}$. Despite a limited
number of data points, there are indications that these spectral
differences also extend to the $10^{35} - 10^{36}$ erg s$^{-1}$
range, but above a luminosity of $10^{35}$ erg
s$^{-1}$ the separation between neutron-star and black-hole systems is
not as clear as below. In addition, the black-hole spectra only become
softer below luminosities of $10^{34}$ erg s$^{-1}$ compared to
$10^{36}$ erg s$^{-1}$ for the neutron-star systems. This observed
difference between the neutron-star binaries and black-hole ones
suggests that the spectral properties (between 0.5--10 keV) at
$10^{34} - 10^{35}$ erg s$^{-1}$ can be used to tentatively determine
the nature of the accretor in unclassified X-ray binaries. More
observations in this luminosity range are needed to determine how
robust this diagnostic tool is and whether or not there are (many)
systems that do not follow the general trend. We discuss our results
in the context of properties of the accretion flow at low luminosities
and we suggest that the observed spectral differences likely arise
from the neutron-star surface becoming dominantly visible in the X-ray
spectra. We also suggest that both the thermal component {\it and} the
non-thermal component might be caused by low-level accretion onto the
neutron-star surface for luminosities below a few times $10^{34}$ erg
s$^{-1}$.

\end{abstract}


\section{Introduction}

In low-mass X-ray binaries (LMXBs), a compact object (a black hole or
a neutron star) is accreting matter from a companion star which
typically has a mass (often significantly) lower than that of the
accretor. The matter is transferred because the donor fills its Roche
lobe. Most LMXBs are so-called X-ray transients: they exhibit sporadic
outbursts during which the X-ray luminosity (0.5-10 keV; from here on
we use this energy range when quoting luminosities, unless otherwise
mentioned) increases up to a few times $10^{38}-10^{39}$ erg s$^{-1}$,
although many systems do not become this bright and some only reach
very faint luminosities of $10^{34}- 10^{36}$ erg s$^{-1}$ \citep[the
so-called very-faint X-ray transients; see, e.g., the discussion
in][]{2006A&A...449.1117W}\footnote{Although
\citet[][]{2006A&A...449.1117W} used the 2--10 keV energy range to
classify the sources, we use in this paper the 0.5--10 keV energy
range in order to study the spectral evolution also below 2 keV.}.
However, most of the time these X-ray transients are in their
quiescent state during which no or hardly any accretion occurs and
consequently their X-ray luminosities are extremely low (typically
$10^{30}-10^{33}$ erg s$^{-1}$).

Ever since LMXBs were first discovered in the late 1960's, they have
been studied intensively using all available X-ray
instruments. Therefore, their observational properties are well
characterized, at least at X-ray luminosities above $\sim10^{36}$ erg
s$^{-1}$. This is mainly due to the limited sensitivity of most X-ray
instruments in orbit in combination with the limited amount of
observing time which can be obtained with high sensitivity instruments
when those X-ray binaries are at very faint
luminosities. Consequently, despite the increasing amount of data over
the last decade, the X-ray behaviour of LMXBs at luminosities of
$10^{34-36}$ erg s$^{-1}$ is not yet understood. We note that for the
purpose of this paper we define systems which are below an X-ray
luminosity of $10^{34}$ erg s$^{-1}$ as quiescent X-ray transients.

The spectral behaviour of LMXBs above $10^{36}$ erg s$^{-1}$ is
relatively well understood \citep[see, e.g., discussions in
][]{2006ARAA..44...49R,2007ApJ...667.1073L}. At lower accretion
luminosities, the phenomenological behaviour is significantly less
clear. Over the last few years, a growing number of very-faint X-ray
binaries have been spectrally studied and the general conclusion is
that the sources can often be satisfactorily described with a simple
power-law model, although more complex models cannot be excluded
because of the limitations imposed by the data quality.  However,
irrespectively of what model is fitted to the spectra, they become
softer with decreasing luminosity. This appears to be true for both
black-hole and neutron-star systems \citep[see, e.g.,][and
references
therein]{2011MNRAS.417..659A,2013MNRAS.428.3083A,2013ApJ...773...59P,2014MNRAS.441.3656R,2015MNRAS.447.1692Y}.

For black-hole systems, this softening is typically explained in the
context of a radiative inefficient accretion flow \citep[e.g., see the
discussion in][]{2013ApJ...773...59P,2014MNRAS.441.3656R}. Similar
explanations have been proposed for the neutron-star systems
\citep[see,
e.g.,][]{2011MNRAS.417..659A,2013MNRAS.434.1586A}. However, when high
quality data are obtained from the neutron-star systems, the spectrum
cannot easily be described with a simple power-law model anymore, but
requires an additional soft component
\citep[e.g.,][]{2013ApJ...767L..31D,2013MNRAS.434.1586A,2013MNRAS.436L..89A}.
Likely this soft emission originates from the surface of the neutron
star, e.g., due to low-level accretion onto the surface which might
produce such type of spectra
\citep[e.g.,][]{1995ApJ...439..849Z,2001A&A...377..955D}. This
additional spectral component complicates the interpretation of the
softening. It is currently unclear if the softening seen in neutron
star systems is due to this soft thermal component becoming dominant
with no change in the power-law component, or if also the power-law
component evolves with decreasing luminosity \citep[see the
discussions in][]{2013MNRAS.434.1586A,2013MNRAS.436L..89A}.

To obtain more insight into these topics, we performed a detailed
literature study about what is currently known about the spectra of
neutron-star X-ray binaries when they have accretion luminosities
between $10^{34}$ and $10^{36}$ erg s$^{-1}$. We will also compare
those systems with the black-hole transients at similar
luminosities. We do not use the symbiotic neutron-star LMXBs
\citep[see][for a compilation of known systems]{2012MNRAS.424.2265L}
in our analysis since in those systems the neutron star accretes from
a stellar wind and not from a disk. In addition, most symbiotic
systems harbour a neutron star with a strong magnetic field (typically
$>10^{12}$ Gauss). Such a strong field will significantly alter the
emerging spectra and therefore those systems cannot be directly
compared with the systems which have a weak or absent magnetic field
($<10^{10}$ Gauss). For the same reason we also do not include the
high-magnetic field (also typically $>10^{12}$ Gauss) neutron-star
systems in which the mass transfer does occur through Roche-lobe
overflow and in which an accretion disk is present (e.g., 4U 1626--67,
4U 1822--37, GRO J1744--28).

\section{Data selection and results \label{selectionresults}}

We searched the literature for publications that describe spectral
results on neutron-star LMXBs (both transients as well as persistent
systems) which are not in quiescence (thus above $10^{34}$ erg
s$^{-1}$; see section~\ref{quiescence} for the justification of this
lower luminosity boundary), but below a X-ray luminosity of a few
times $10^{36}$ erg s$^{-1}$.  Since it is unclear if a dynamically
important neutron-star magnetic field could alter the observed X-ray
spectra, we initially only use the non-pulsating systems\footnote{Aql
X-1 has shown a very brief episode during which it exhibited X-ray
pulsations \citep[][]{2008ApJ...674L..41C}, but the vast majority of
its time it does not exhibit such oscillations and its general X-ray
spectral and timing behaviour resembles other non-pulsating
LMXBs. Therefore, we included this source in our non-pulsating
sample.}. In section~\ref{amxps} we compare the non-pulsating systems
with the accreting millisecond X-ray pulsars to investigate if indeed
magnetic field effects could be observed in the X-ray spectra in
neutron-star LMXBs.

We limited ourselves to those publications which reported on
observations that covered the 0.5--10 keV energy range, so that
accurate measurements could be obtained for the spectral shape below
$\sim$2 keV. In addition, those systems that have a column density of
$N_{\rm H} >5 \times 10^{22}$ cm$^{-2}$ were excluded from our source
selection because their spectra below 2 keV cannot be accurately
modelled. Extrapolation from the 2--10 keV fit results down to 0.5 keV
can result in significant systematic errors. This criterion excludes
most sources close to the Galactic centre \citep[e.g., those reported
on
by][]{2002ApJ...579..422W,2003ApJ...598..474M,2005ApJ...622L.113M,2005MNRAS.357.1211S,2005AA...430L...9P,2007AA...468L..17D,2009AA...495..547D,2010AA...524A..69D,2013IAUS..290..113D,2011MNRAS.414L.104D,2012AA...545A..49D}.
In section~\ref{section_nh} we will briefly discuss those sources and
how including them in our source sample would affect our conclusions.

\begin{figure*}
 \begin{center}
\includegraphics[width=2\columnwidth]{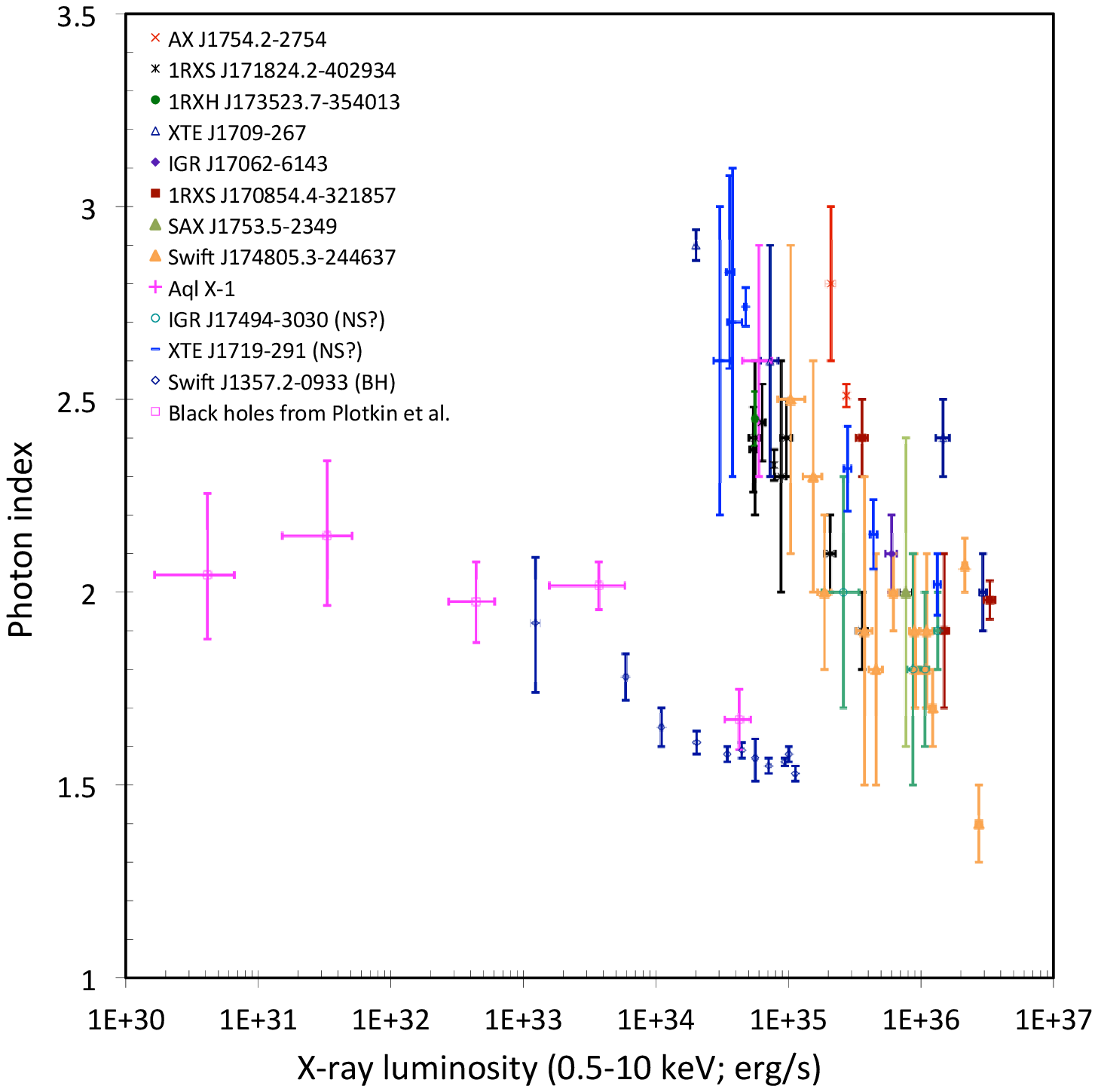}
    \end{center}
 \caption[]{The photon index versus the 0.5-10 keV X-ray luminosity for the neutron-star X-ray binaries used in this paper (Tab.~\ref{sources}) as well as for the black-hole system Swift J1357.2-0933 \citep{2013MNRAS.428.3083A} and the average of the black-hole  points from \citet{2013ApJ...773...59P}. (A colour version of this figure is available in the online version of this paper.)}
 \label{photonindex}
\end{figure*}

In most publications the spectral results were reported when using a
simple absorbed power-law model, but often also a two-component model
was used: a thermal component at low energies and a power-law
component at high energies
\citep[e.g.,][]{2004MNRAS.354..666J,2010ApJ...714..270F,2013ApJ...767L..31D,2013MNRAS.434.1586A,2008ApJ...684L..99C,2014MNRAS.441.1984C}.
Sadly, no uniform model was used \citep[e.g., for the thermal
component either a black-body model or a neutron-star atmosphere model
was used; e.g.,
][]{2004MNRAS.354..666J,2013MNRAS.434.1586A,2013MNRAS.436L..89A}
making global comparisons between sources difficult. Since for most
sources the results of a single power-law model were reported (which
typically resulted in acceptable fits), we will focus on that model in
this paper. When a two-component model was used and no results were
reported on a single power-law model \citep[e.g.,
in][]{2010ApJ...714..270F,2011ApJ...736..162F,2008ApJ...684L..99C,2014MNRAS.441.1984C},
we did not include those points in our selection. Photon indices
($\Gamma$) obtained from fitting a two-component model are typically
very different from those obtained with a single power-law model. This
is clear from the papers where both types of fits are reported
\citep[e.g., ][]{2013MNRAS.434.1586A}.

\begin{figure}
 \begin{center}
\includegraphics[width=1.25\columnwidth]{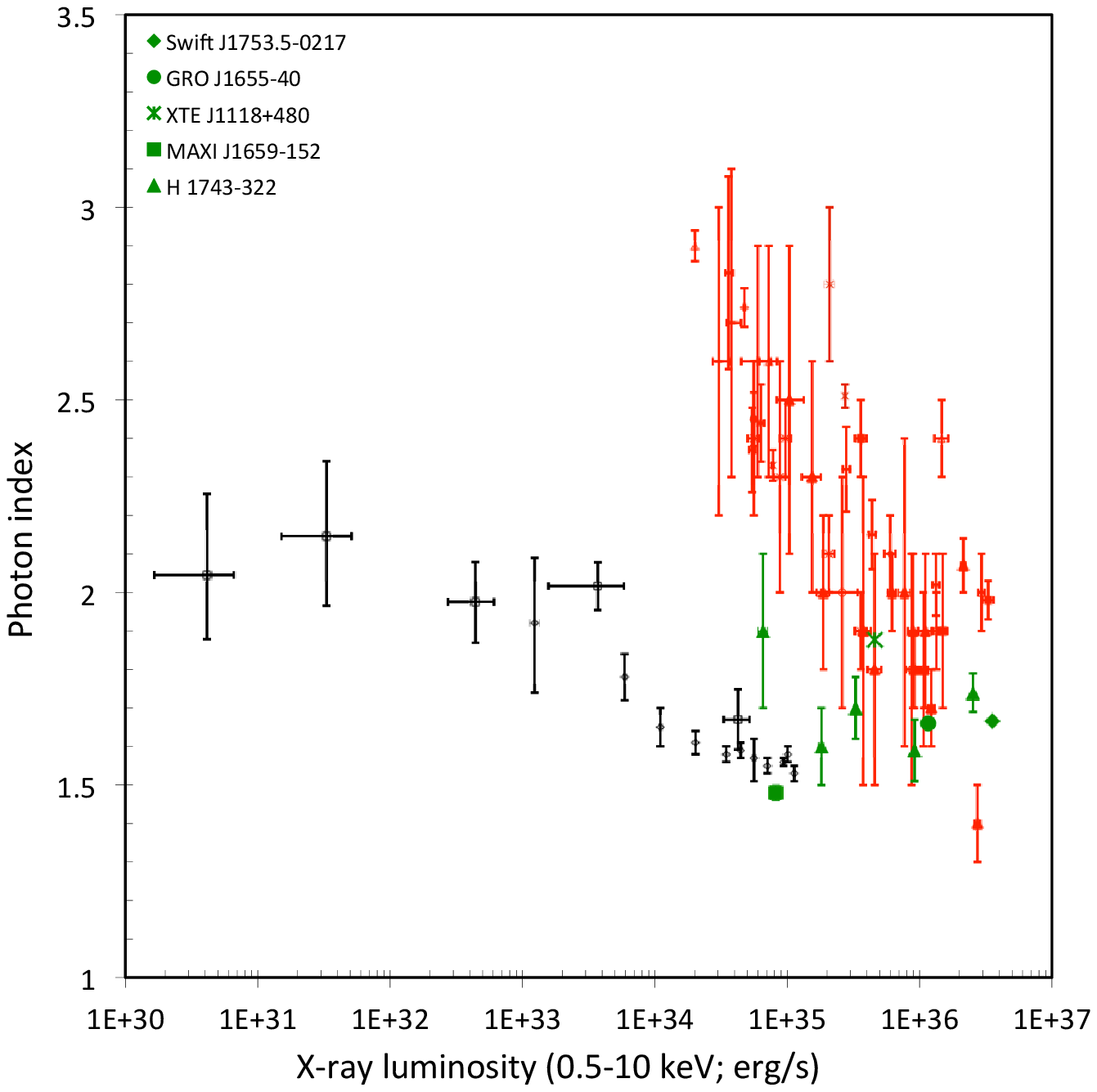}
    \end{center}
 \caption[]{Adapted from Figure~\ref{photonindex}. The red points are the neutron-star systems and the black points the black-hole transients. In addition, the figure now includes several additional black-hole X-ray transients with luminosities in the range $10^{34}$ to $10^{36}$ erg s$^{-1}$ (the green points). (A colour version of this figure is available in the online version of this paper.)}
 \label{photonindexBH}
\end{figure}

In some publications where the sources were fitted with a single
power-law model, the errors on the obtained photon indices were so
large that no significant conclusions can be obtained from those data
points. Therefore, we do not use those data points for which the
photon indices had an error larger than 0.5. This excludes the results
in several publications such as those reported for SAX J1750.8--2900
\citep[][]{2012ApJ...749..111L,2013MNRAS.434.1599W,2015ApJ...801...10A},
SAX J1828.5--1037 \citep[][]{2008ATel.1831....1D,2009ApJ...699.1144C},
or Swift J1749.4-2807
\citep[][]{2009MNRAS.393..126W,2009ApJ...699.1144C}. In addition, this
excludes some of the data points for some of the sources we do use in
this paper \citep[e.g., some data points of Swift
J174805.3--244637;][]{2014ApJ...780..127B}.

\citet[][]{2012ApJ...759....8D} reported on monitoring observations
using {\it Swift} of two newly discovered neutron-star X-ray
transients: Swift J185003.2--005627 and Swift J1922.7--1716.  Also in
that paper the reported results were based mostly on a two component
model to fit the source spectra \citep[see also][for Swift
J1922.7--1716]{2006A&A...456L...5F} but they also listed the results
obtained using a single power-law model. However, those results were
obtained from a spectral fit to the combined data set including a
range of X-ray luminosities so that any possible evolution of the
X-ray spectral shape with luminosity is averaged out. Therefore, we
also do not include these results in our analysis. We note that for
Swift J18500.3--005627, \citet[][]{2012ApJ...759....8D} found that
(when studying the X-ray colours) the source softened when the X-ray
luminosity decreases. This would be consistent with the general trend
for neutron-star X-ray binaries we report on below.

Our resulting sample of neutron-star X-ray binaries is listed in
Table~\ref{sources}. We have included all sources in our sample for
which we found results in the published literature. However, it is
conceivable that we did not find all the relevant publications and
that therefore some sources could have been missed. However, we feel
that that would only be a very low number of sources and therefore the
sources used in our paper should serve as a representative sample of
the whole population. Among our eleven sources, three are persistently
accreting at $10^{34} - 10^{-35}$ erg s$^{-1}$ and eight are X-ray
transients (see Tab.~\ref{sources}). Nine sources are confirmed
neutron-star binaries since type-I X-ray bursts have been seen, and
two are strong neutron-star candidates (see section
\ref{NS_candidates}). In Figure~\ref{photonindex} we plot the photon
indices of those eleven sources versus their 0.5-10 keV X-ray
luminosities. Although there is a large scatter in the points, the
neutron-star systems follow a clear trend: on average the photon index
is between 1.5 and 2 for X-ray luminosities of $\sim$10$^{36}$ erg
s$^{-1}$, but it increases to 2--3 (i.e., the overall X-ray spectrum
becomes softer) when the X-ray luminosity decreases. This was observed
already from some individual sources that showed a large dynamic
range in luminosity \citep[see,
e.g.,][]{2011MNRAS.417..659A,2014ApJ...780..127B,2014MNRAS.438..251L}
but from Figure~\ref{photonindex} it can now be seen that all
neutron-star systems in our sample follow in general a similar
behaviour.

\begin{figure}
 \begin{center}
\includegraphics[width=\columnwidth]{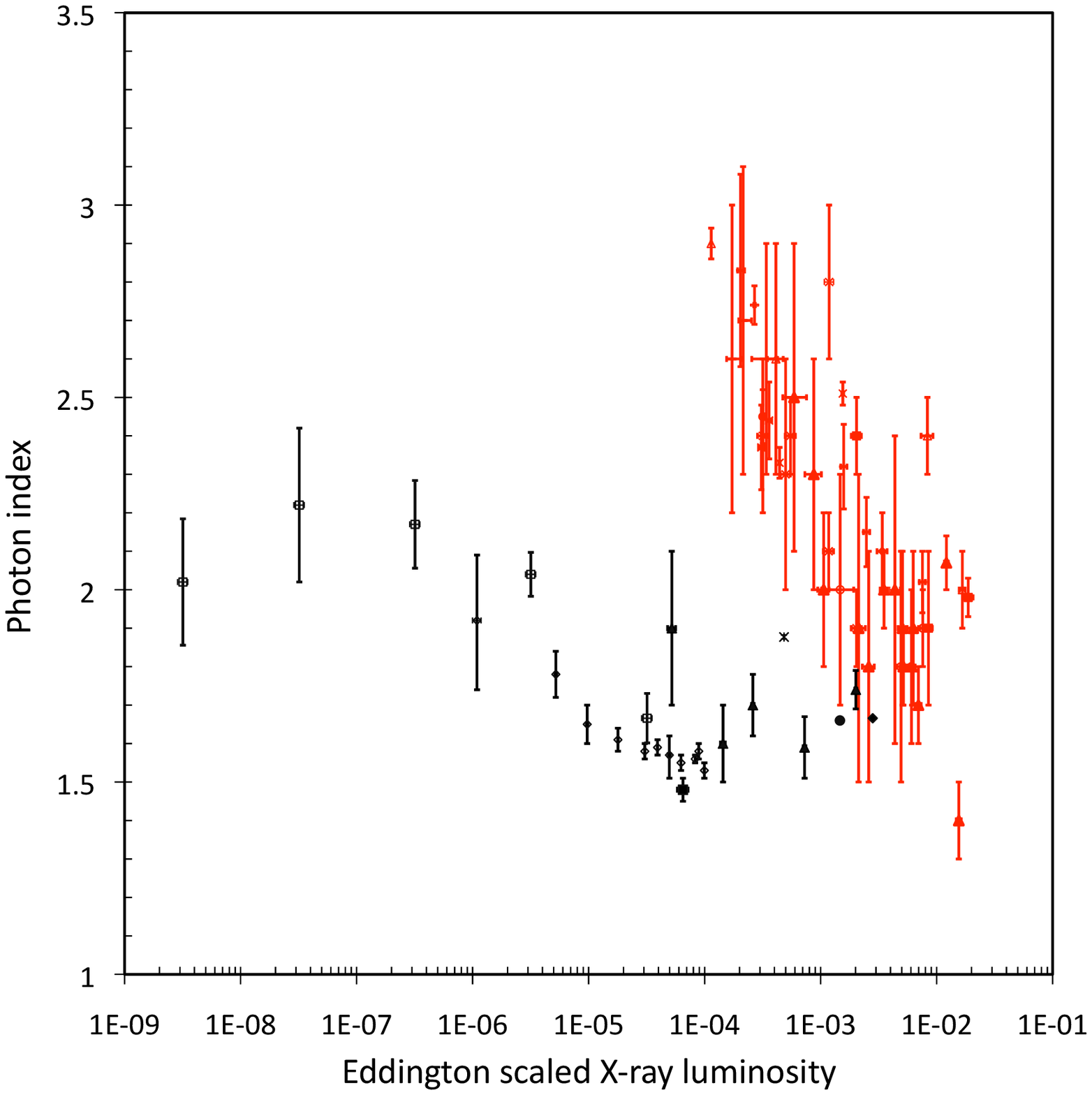}
    \end{center}
 \caption[]{Adapted from Figure~\ref{photonindexBH} but scaled to the Eddington X-ray luminosity (the black points are the black-hole systems; the red points the neutron-star ones; a colour version of this figure is available in the online version of this paper).}
 \label{photonindexEdd}
\end{figure}

We performed several correlation tests to determine if indeed the
photon index is correlated with the luminosity. A Pearson test,
Spearman test, and a Kendall test resulted in correlation indices of
$-0.81$, $-0.77$, and $-0.60$, respectively, corresponding to a
probability of $1.78\times10^{-11}$, $4.83\times10^{-10}$, and
$6.84\times10^{-9}$, respectively, that both parameters are
uncorrelated. Therefore, we consider these results a support for our
conclusions that the photon index is indeed anti-correlated with the
luminosity. We fitted a power law of the shape $\Gamma = a \log L_{\rm
X} + b$ to the data and we obtained the following parameters: $a =
-0.42 \pm 0.03$ and $b = 17.3 \pm 1.1$ (90\% confidence
levels). To quantify the scatter around this correlation we
calculated the root-mean-square of deviations from the corrlation and
found a value of 0.28.

\subsection{Comparison with black-hole X-ray binaries \label{KSintro}}

To compare our result obtained for the neutron-star LMXBs with
black-hole systems, we used the data reported by
\citet{2013ApJ...773...59P} who discussed the behaviour of black-hole
X-ray transients as they decayed down to quiescence. We applied
the same selection criteria as we used for our neutron-star sample to
the invidual black-hole points reported by
\citet[][their Table 3 and 5]{2013ApJ...773...59P}. We averaged the photon indices and
calculated the corresponding errors. We averaged the luminosities and
as errors we used the standard deviation on those luminosity points.
In addition, we used the points of Swift J1357.2-0933 from \citet[][we
used the averaged points used to create their Figure 1, right panels;
the values of those points are not listed in this reference, so we
list them in Table \ref{tabBH}]{2013MNRAS.428.3083A}. Clearly, the
black-hole points are significantly offset from the neutron-star
points in Figure~\ref{photonindex}.

We searched the literature for more black-hole transients with
luminosities between $10^{35}$ and $10^{37}$ erg s$^{-1}$ but not many
could be found. The ones we found are listed in Table~\ref{tabBH} and
shown in Figure.~\ref{photonindexBH}.  Even though in the luminosity
range of $10^{35} - 10^{36}$ erg s$^{-1}$ the black-hole systems
appear to be on average harder than the neutron-star binaries, there
is a significant overlap of the data points (especially above $\sim 5
\times 10^{35}$ erg s$^{-1}$). We only have a few black-hole systems
in this luminosity range making final conclusions on the average
hardness of those systems in this range difficult. In any case, it
seems that above $10^{35}$ erg s$^{-1}$ the neutron-star binaries are
not as clearly separated from the black-hole transients as at
luminosities below $10^{35}$ erg s$^{-1}$ (even more so when
also including the accreting millisecond X-ray pulsars, which are
discussed in section \ref{amxps}).

We used a 2D Kolmogorov-Smirnov (KS) test
\citep[see][]{1987MNRAS.225..155F} on the neutron-star and black-hole
data points to quantify the probability that the two samples are drawn
from two different distributions, therefore confirming that they
behave differently.  However, the KS test does not account directly for the
errors in the measurements of photon index and luminosity. In order to
account for such errors we generated $10^5$ synthetic samples
randomly shifting each source around the reported values. We assumed
a Gaussian distribution with standard deviations conservatively given
by the reported errors on the measurements. First, we compared the
full data sets of neutron stars and black hole; the black hole sample
included the (not averaged) values of \cite{2013ApJ...773...59P}, as well
as the other black hole data (see Fig.~\ref{photonindexPlotkin} and section
\ref{individual}). Assuming a null hypothesis that the two distributions are the
same, we obtained a 90 $\%$ confidence interval for the probability of
$2.4 \times 10^{-14}$ - $3.3 \times 10^{-11}$. This result strongly
supports our idea that the two data sets are drawn from different
distributions. However, many black-hole data points have luminosities
below $10^{34}$ erg s$^{-1}$; that may influence the results of the test,
since no neutron-star systems were include below that threshold. To
avoid biasing the results, we repeated the test keeping only the
black-hole data points with luminosities above $10^{34}$ erg/s. In
this case the 90 $\%$ confidence interval for the probability that the
two sets come from the same distribution is higher, but still it is
very low, being only $2.8 \times 10^{-7}$ - $2.7\times
10^{-5}$. Despite the low number of data points, we conclude that the
2D KS test supports our idea that the two kind of systems behave
markedly differently.

\citet{2013ApJ...773...59P} used the luminosities
scaled to the Eddington luminosity when discussing the black-hole
systems. For quiescent X-ray transients there are physical reasons for
doing this \citep[see, e.g.,][]{1999ApJ...520..276M} but it is unclear
if those reasons also apply when the sources are not in
quiescence. However, for completeness, in Figure~\ref{photonindexEdd}
we replotted our data set with the X-ray luminosity in Eddington
units. We used the Eddington scaled values from
\cite{2013ApJ...773...59P} as presented in their Figure 4b. We
converted the points of Swift J1357.2--09333
\citep[from][]{2013MNRAS.428.3083A} using a black-hole mass of 9
$M_{\odot}$, and we used the same black-hole masses used by
\citet[][]{2013ApJ...773...59P} for the other black holes \citep[except for
Swift J1753.5--0127 for which we assumed 10 $M_\odot$ since this
source was not included in the study of][]{2013ApJ...773...59P}.  We converted the neutron-star
values using a neutron-star mass of 1.4 $M_{\odot}$ (resulting in an
Eddington luminosity of $1.8 \times 10^{38}$ erg s$^{-1}$). It is
clear that when using Eddington scaled luminosities, the difference
between the neutron-star systems and the black-hole systems becomes
even more pronounced. We note that when using the Eddington scaled
luminosities, an additional uncertainty is introduced because of the
uncertainties in the mass of the accretors. Most black-hole masses are
poorly constrained and also the neutron-star masses could display a
range (e.g., from 1.2 up to 2 $M_\odot$).  Therefore, from here on, we
only show the X-ray luminosities in our figures and not the Eddington
scaled one since this allows for direct comparison of different source
types without assuming anything about the nature and mass of the
accretor.

\subsubsection{Individual black-hole systems of Plotkin et al. (2013) \label{individual}}

\begin{figure}
 \begin{center}
\includegraphics[width=\columnwidth]{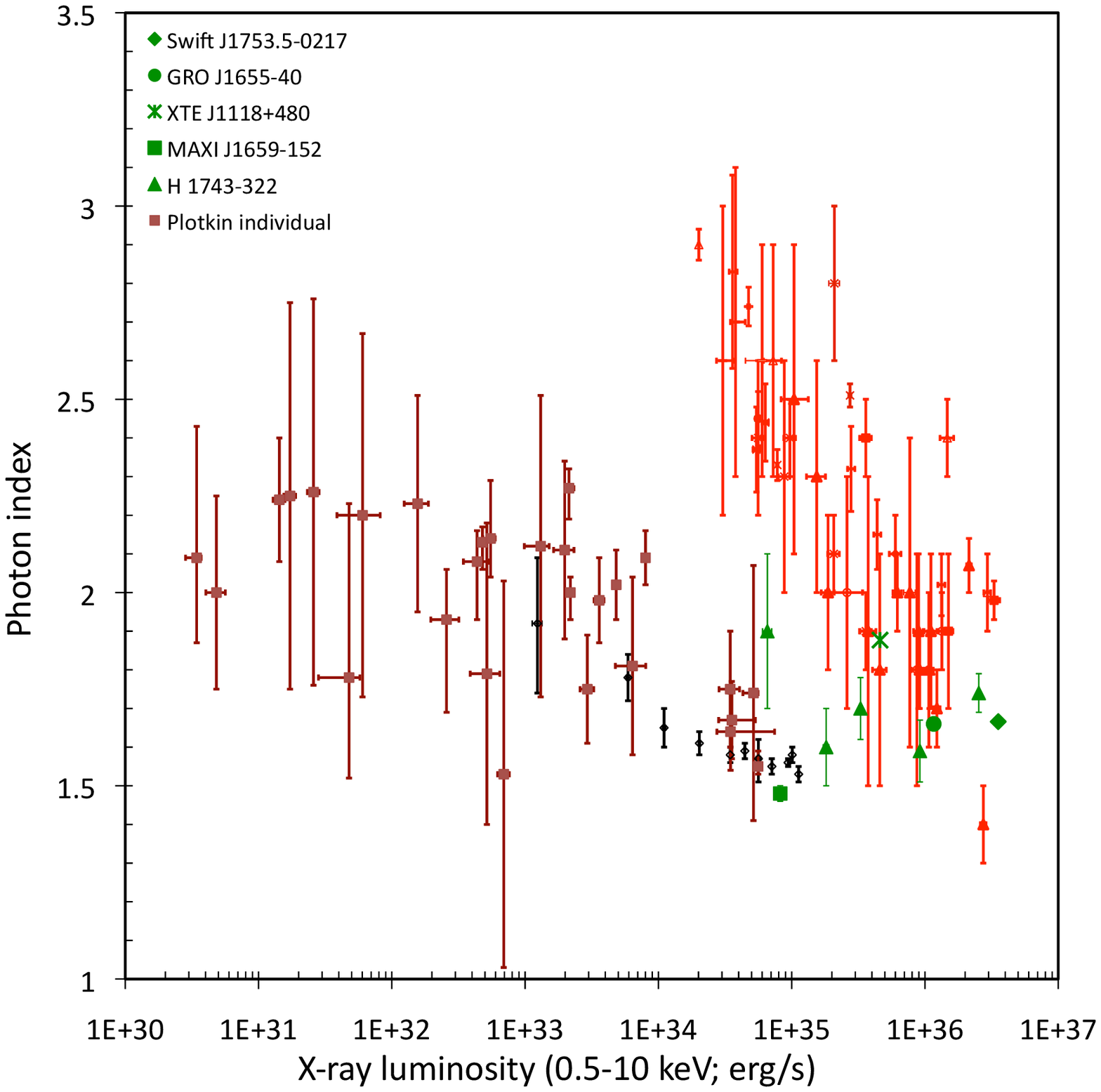}
    \end{center}
 \caption[]{Similar to Figure~\ref{photonindexBH} but with the black-hole systems from \citet[][]{2013ApJ...773...59P} plotted as individual points.  (A colour version of this figure is available in the online version of this paper.)}
 \label{photonindexPlotkin}
\end{figure}

In Figure~\ref{photonindexBH} we showed the average data for the
black-hole X-ray transients from \citet[][]{2013ApJ...773...59P}. Such
averaging might make potential outliers in the black-hole sample
invisible. It could be possible that some systems have X-ray spectra
considerably softer than the average (and thus possibly more
consistent with the neutron-star systems than on average). Therefore,
in Figure~\ref{photonindexPlotkin} we show the invididual black-hole
systems.  Clearly this figure shows that, despite the larger errors
and the scatter in the data points, none of the black-hole systems
lies significantly above the average track. Therefore, we are
convinced that using the average data points does not introduce
biases.

\subsection{Comparison with accreting millisecond X-ray pulsars \label{amxps}}

\begin{figure}
 \begin{center}
\includegraphics[width=1.2\columnwidth]{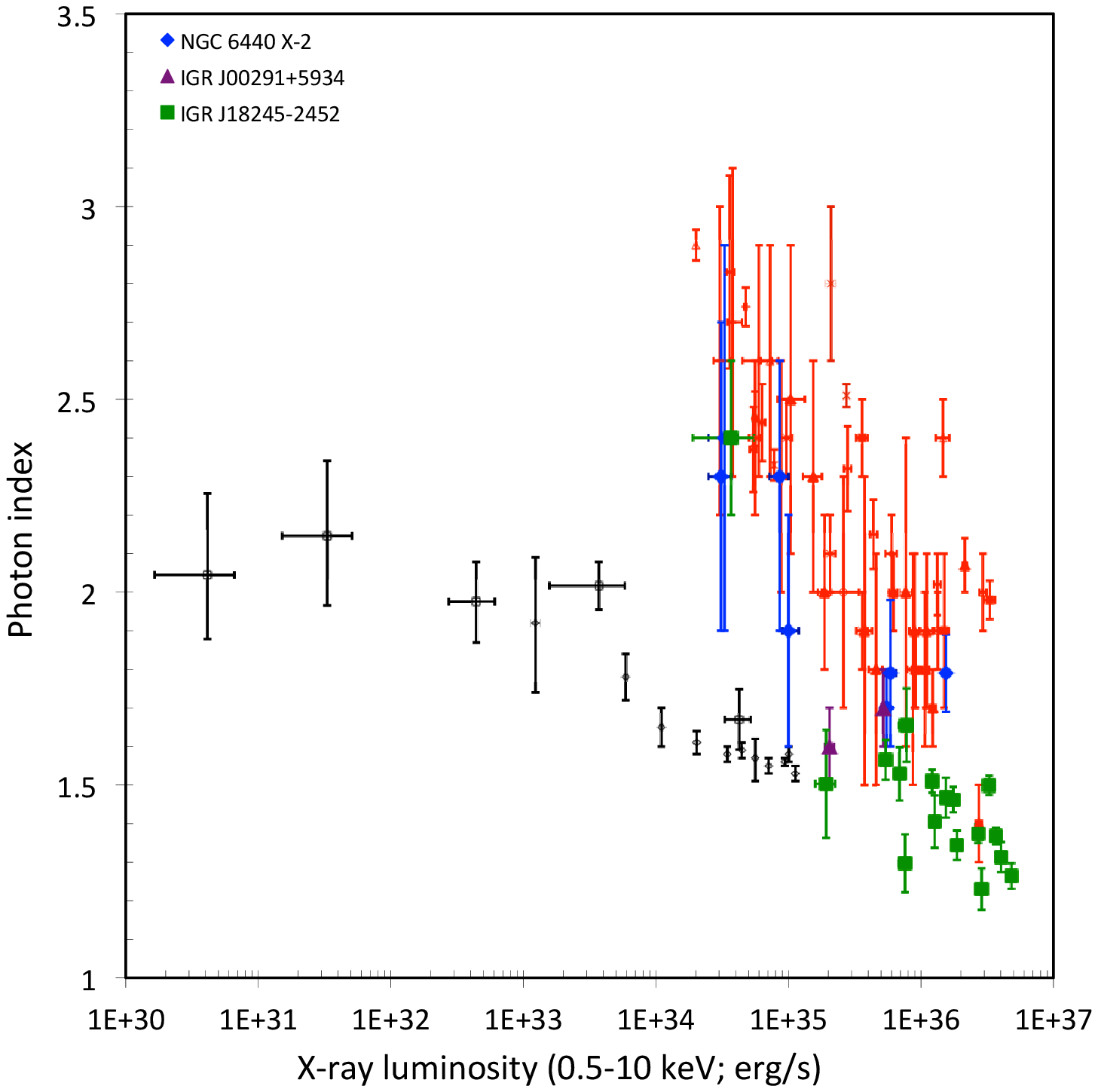}
    \end{center}
 \caption[]{Adapted from Figure~\ref{photonindex}. The red points are
 the neutron-star systems and the black ones the black-hole
 binaries. Also included are the data points (green) for the transient
 and AMXP IGR J18245--2452 which is located in the globular cluster
 M28 and two other AMXPs (NGC 6440 X-2:
 blue; IGR J00291+5934: purple; a colour version of this figure is
 available in the online version of this paper).}
 \label{photonindexAMXPs}
\end{figure}

In Figure~\ref{photonindexAMXPs} we compare the three accreting
millisecond X-ray pulsars for which we found appropriate data points
in the literature with the non-pulsating systems. From this figure it
appears that the pulsating sytems are harder (at the same
luminosities) than the non-pulsating systems. In particular the X-ray
transient IGR J18245--2452 located in the globular cluster M28 is very
hard \citep[i.e., above $10^{36}$ erg
s$^{-1}$;][]{2014MNRAS.438..251L}, even harder than the black-hole
systems\footnote{The small errors on the column densities for this
source, as reported in Table~\ref{sourcesAMXP}, strongly indicated
that an underestimation of the column density could not have resulted
in the very low photon indices measured. In addition, for the quoted
luminosities the unresolved faint-source background in M28 does not
affect the obtained spectral parameters \citep{2014MNRAS.438..251L}}. This is contrary to the generally believed idea that black-hole
systems are typically harder than the neutron-star ones. However, this
source is on of the very few neutron-star X-ray transient that
transits between an accreting millisecond X-ray pulsar when in
outburst and a millisecond radio pulsar when in quiescence
\citep[][]{2013Natur.501..517P}. It is quite possible that its very
hard spectra might be related to this unique character (although this
would not remove the fact that some neutron-star X-ray transients can
be harder than the black-hole systems). Moreover, the other AMXPs do
not have such hard spectra when in outburst, although maybe the
spectra might still be (slightly) harder than the non-pulsating
systems.

We used again a 2D KS test to check whether the AMXP data were
consistent with the data from the non-pulsating neutron-star systems
applying the same method of section \ref{KSintro}. We found a 90 $\%$
confidence interval for the probability that they are the same of
$1.2 \times 10^{-6}$ - $3.5 \times 10^{-4}$. However, since we only
have three systems and IGR J18245--2452 might dominate the
distribution for the AMXPs, we redid the KS test but now with this
source removed. In that case, we found a probability interval of $4.8
\times 10^{-3}$ - $3.6 \times 10^{-1}$ that the AMXPs have the same
distribution as the non-pulsating systems. This implies that we cannot
draw any strong conclusions, especially given the fact that our AMXP
sample is very limited. Therefore we cannot state conclusively that
the presence of a dynamically important magnetic field alters the
X-ray spectra of neutron-star LMXBs.

\subsection{Below $\mathbf{10^{34}}$ erg s$\mathbf{^{-1}}$: quiescence \label{quiescence}}

The black-hole systems we consider go to significantly lower
luminosities than the neutron-star systems because for the black holes
we did not apply the $10^{34}$ erg s$^{-1}$ cut-off criterion as we
did for the neutron-star systems. When we include similarly low
luminosities for the neutron-star systems, the situation becomes very
complex: some systems are totally thermally dominated resulting in
very large photon indices \citep[3--5;
e.g.,][]{2001ApJ...559.1054R,2002ApJ...580..413R,2004ApJ...610..933T}
but others are power-law dominated with photon indices as low as
$\sim$1.0-1.8
\citep[e.g.,][]{2002ApJ...575L..15C,2005ApJ...618..883W,2007ApJ...660.1424H,2009ApJ...691.1035H,2012ApJ...756..148D}. The
strong thermal emission could be due to residual accretion on the
surface, but likely in many systems it is due to the cooling emission
from the neutron star that has been heated during the outburst
\citep[e.g.,][]{1998ApJ...504L..95B,1999ApJ...514..945R,2001ApJ...551..921R,2001ApJ...559.1054R,2000A&A...358..583C}. The
power-law dominated spectra for quiescent neutron-star transients are
not understood \citep[see, e.g., the discussion
in][]{1998A&ARv...8..279C} but the power-law component could be
related to the fact that the neutron star has a magnetosphere
\citep[see][for a recent discussion]{2012ApJ...756..148D}. Because of
these complications and the large range of photon indices in the
neutron-star systems, we do not plot the neutron-star data for
luminosities below $10^{34}$ erg s$^{-1}$. For black-hole systems, the
situation is much cleaner since any effect on the spectra by a solid
surface or a magnetic field is absent and all the X-rays should be due
to some form of low-level accretion. This is reflected in the rather
homogenous behaviour of the black-hole systems in their quiescent
state \citep[see][for a detailed discussion]{2013ApJ...773...59P}. It
is interesting to note that at quiescent luminosities some
neutron-star systems have significantly harder X-ray spectra than the
black-hole transients \citep[e.g., SAX J1808.4--3658, EXO
1745--248][]{2002ApJ...575L..15C,2009ApJ...691.1035H,2012MNRAS.422..581D},
although most of these hard systems (except EXO 1745--248) are
accreting millisecond X-ray pulsars and their hard quiescent spectra
might be related to the presence of a dynamically important magnetic
field \cite[see discussion in ][]{2012ApJ...756..148D}.

\section{Uncertainties}

The neutron-star points in Figure~\ref{photonindex} display
significant scatter, which could be intrinsic due to source
differences. If true, then this would imply that although all sources
become significantly softer at low luminosities, they might not all
follow exactly the same uniform relation with decreasing
luminosity. However, several types of systematic uncertainties are
likely to be present in the data which would also give rise to
significant scatter. Therefore a uniform relationship for all sources
cannot be excluded at present. The main possible contributors to the
scatter are discussed below. However, despite those uncertainties, we
are convinced that the difference between the neutron-star systems and
the black-hole ones in the luminosity range between $10^{34}$ and
$10^{35}$ erg s$^{-1}$ is robust.

\subsection{The distance \label{distance}} 

For the neutron-star systems the distance is usually determined from
the observation of type-I X-ray bursts, except for XTE J1719-291 and
IGR J17494-3030 which so far have not displayed type-I X-ray bursts
(we used for both sources a distance of 8 kpc because of their
proximity to the Galactic centre; Table~\ref{sources}). Depending on
the assumed burst properties (e.g., what type of fuel is burning on
the surface and in which type of environment such as Hydrogen rich or
Hydrogen poor) the obtained distances can vary by 50\% or more. E.g.,
for AX J1754.2-2754 a distance of 6.6 or 9.2 kpc was obtained by
\citet{2007AstL...33..807C}. In Figure~\ref{photonindex} we used a
distance of 9.2 kpc, but if the distance would be 6.6 kpc, the X-ray
luminosity would decrease by a factor of $\sim$2. This would shift the
source more in line with the other sources because now it has
relatively soft photon indices for its luminosity compared to the
majority of sources (Fig.~\ref{photonindex}). However, similar
uncertainties exist in the distance measurements of the other sources,
leading up to uncertainties in the X-ray luminosity of a factor 2 or
3. However, even if all neutron-star sources would shift down by this
factor, they would still not be consistent with the black-hole
ones. Moreover, for several sources we might have actually
underestimated the distances instead of overestimating them, which
would move some sources to higher luminosities and not lower
luminosities, strengthening the difference between black-hole and
neutron-star LMXBs.

We note that similar distance uncertainties exist in the black-hole
data used by \citet{2013ApJ...773...59P}.  In addition, for Swift
J1357.2--0933 we used a distance of 1.5 kpc. However, recently it has
been argued that the distance for this source could range from 0.5 kpc
up to 6.3 kpc \citep[][]{2013MNRAS.434.2696S}. Using this distance
range would move the data down by a factor of up to 9, making the
source exceptionally faint in outburst as well as in quiescence
\citep[][]{2014arXiv1404.2134A}, or up by a factor of up to 18, which
still would make the system not consistent with the neutron-star
systems.

\subsection{Pile-up effects}

Despite the low X-ray luminosities, the source fluxes can be still
high enough to produce significant pile-up during some of the
observations, especially if taken with {\it Chandra} in imaging mode
\citep[e.g., ][]{2005AA...440..287I}.  The effect would be that the
spectra would appear artificially harder than they really are.  Since
we found that the brightest sources have the hardest spectra, the
question arises whether this could be due to pile-up effects. However,
in our selection we did not include data points which are
significantly affected by pile-up. Additionally, a lot of data were
obtained using {\it XMM-Newton} or {\it Swift} which are less
sensitive to pile-up and often the high-time resolution mode was used
when the sources were relatively bright. Although we cannot exclude
that some data points are still affected by a small amount of pile-up,
most sources did not show evidence for pile-up \citep[see, e.g.,
][]{2013MNRAS.434.1586A} so likely the effect of pile-up will be small
and cannot explain the overall correlation we found.

\begin{figure}
 \begin{center}
\includegraphics[width=1.2\columnwidth]{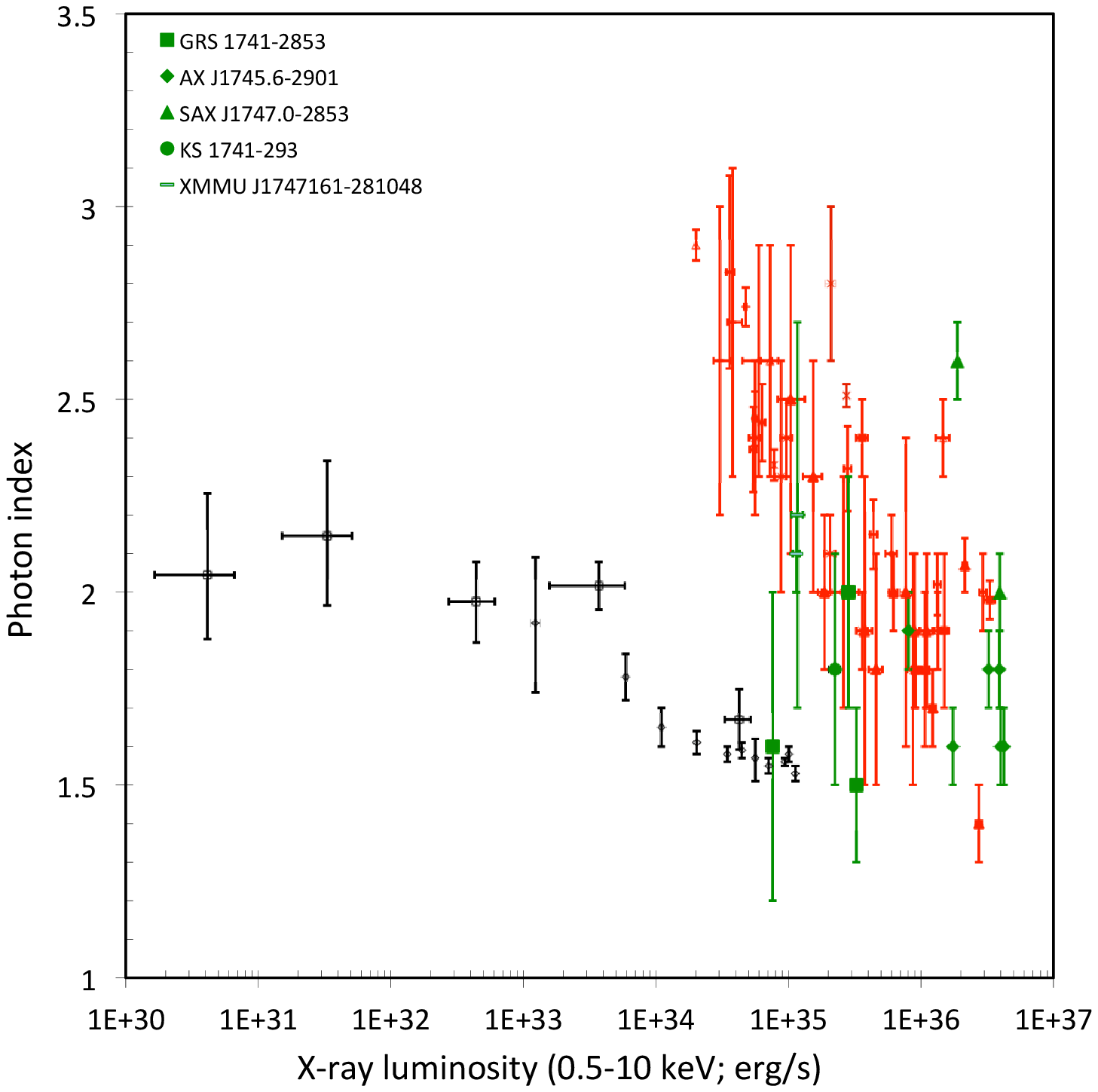}
    \end{center}
 \caption[]{Adapted from Figure~\ref{photonindex} but with the black holes displayed as the black points and the neutron stars as the red points. Also shown are several neutron-star transients which have very high column densities (the green points). (A colour version of this figure is available in the online version of this paper.)}
 \label{photonindexhighNh}
\end{figure}

\subsection{Column density \label{section_nh}} 

Despite the fact that we do not use the neutron-star systems which
have a $N_{\rm H} > 5 \times 10^{22}$ cm$^{-2}$, the range in $N_{\rm
H}$ is still large (ranging from $0.2 \times10^{22}$ cm$^{-2}$ to
$\sim 3 \times 10^{22}$ cm$^{-2}$; Tab.~\ref{sources}). This range
might still produce a systematic effect on the photon indices since it
is well known that when fitting a power-law model to relatively poor
quality data, a strong correlation will be found between the photon
index and the column density \citep[see, e.g., the detailed study
performed on faint black-hole systems
by][]{2013ApJ...773...59P}. However, we would like to stress
here that for the few sources in our sample that have a high dynamic
range in luminosities, that they individually showed this softening
with decreasing luminosities \citep[see,
e.g.,][]{2011MNRAS.417..659A,2014ApJ...780..127B,2014MNRAS.438..251L}. For
those sources the systematic effects of using an incorrect column
density should be minimal.

For the complete sample, any systematic effect from obtaining an
incorrect column density should be reflected in the errors on the
photon indices. However, on several occasions the column density was
frozen to an assumed value (e.g., the expected Galactic column density
towards the source or the value obtained during higher quality data
sets from the sources). If the wrong value is assumed, this would skew
the photon index to higher or lower values than the true one and
introduce scatter in the diagram, although it is currently not clear
how much scatter this would produce. A detailed study of this is
beyond the scope of this paper.

When including those sources that have very high column densities ($>5
\times 10^{22}$ cm$^{-2}$), most of those sources are consistent with
the other neutron-star data points (Figure~\ref{photonindexhighNh})
although there seems to be some indications that at the same
luminosity (below $10^{36}$ erg s$^{-1}$) the highly absorbed sources
have a smaller photon index compared with that observed for the
low-N$_{\rm H}$ sources.  We performed a 2D KS test between the data
points for systems with a low column density and those that have a
high column density. The result is a 90 $\%$ confidence interval for
the probability that they are drawn from the same distribution of
$8.7\times 10^{-4}$ - $6.4 \times 10^{-2}$. Although this
result might indicate that their distributions could be different, the
low column density data set goes to significantly lower luminosities
than the high column density data set. Therefore, we performed also a
2D KS test between the two data sets but limiting the low column
density data points to the same luminosity range as the high column
density ones (i.e., from luminosities $>7 \times 10^{34}$ erg
s$^{-1}$). The resulting 90$\%$ confidence interval increases to
$4.5\times 10^{-3} - 1.7 \times 10^{-1}$, indicating that it is less likely
that the two data sets are drawn from different distributions.  
However, given the limited amount of data points, we cannot make a
strong conclusion.  Moreover, as we will argue in section
~\ref{origin}, the softening is likely due to the neutron-star surface
becoming prominently visible at low energies. Therefore the softening
concentrates at low energies in the X-ray spectra where also the
absorption effects the spectra the most. As a consequence the highly
absorbed sources will show less apparent softening than the less
absorbed systems and any possible difference is therefore likely not
physical.

\subsection{Calculating the 0.5-10 keV luminosities and its errors \label{errorsLx}}

Not all publications we used quoted the luminosities or unabsorbed
fluxes in the 0.5--10 keV band. Some publications quoted only the 2-10
keV or the 0.3-10 keV luminosities. We used
WebPIMMS\footnote{Available at
http://heasarc.gsfc.nasa.gov/Tools/w3pimms.html} to calculate the
luminosities in the 0.5-10 keV energy range, but this introduces
additional uncertainties since extrapolation outside the specified
energy range may not be valid.  In addition, in some publications the
luminosities or fluxes are reported but not their uncertainties (see
Tab.~\ref{sources}); in such cases we assumed an error of 10\% on the
fluxes, although this is likely an underestimation of the true
errors. However, both types of uncertainties will likely have only a
small effect on the observed correlation seen in
Figure~\ref{photonindex} because the scatter in X-ray luminosity is
dominated by the uncertainties in the distances which typically result
in luminosity uncertainties of a factor of a few (see discussion in
section~\ref{distance}).

\subsection{Complex spectral shape \label{section_thermal}}

When the data have low statistics, a single power-law model can
adequately fit the X-ray spectra, but for high quality data a single
power-law model often does not provide a good fit.  Adding a second
(often assumed to be thermal) component is required. Moreover, in the
discussion section we will argue that in most, if not all,
neutron-star systems, a soft thermal component might be present at the
lowest luminosities even if it cannot be detected due to the
limitation of the data. To obtain a homogenous data sample, we have
still included the results obtained using a single power-law model of
those sources for which indeed such a soft component is clearly needed
to fit the spectra accurately. By doing so possible systematic effects
might have been introduced.  However, likely they are not very
significant because when low (e.g., {\it Swift}) and high (e.g., {\it
XMM-Newton}) quality data are available for the same source at roughly
the same X-ray luminosities, the photon indices (when fitting a single
power-law model) obtained from the different data sets are fully
consistent with each other \citep[see, e.g.,][]{2011MNRAS.417..659A}.

\subsection{The effect of individual sources}

Since we are interested in the average behaviour of our sample, it is
possible that individual sources might not follow exactly the
correlation found for the total sample of sources. For example, the
black-hole point in Figure ~\ref{photonindexBH} that has luminosity of
$\sim 5 \times 10^{35}$ ergs $^{-1}$ but a relatively large photon
index of $\sim1.88$ is the only available point of XTE J1118+480
\citep{2010MNRAS.402..836R}. Excluding this point would cause the
black-hole points in Figure ~\ref{photonindexBH} to overlap less with
the neutron-star data for luminosities $> 10^{35}$ erg
s$^{-1}$. Therefore, excluding this point would strengthen the
conclusion that neutron-star systems are softer than black-hole
transients. And indeed arguments can be put forward to exclude this
point because it was obtained from high quality data taken with the
{\it Chandra}/LETG instrument which is particularly sensitive at low
energies. Furthermore, the X-ray spectra were only fitted up to
$\sim$7 keV (due to the instrument response) making the lower energies
more dominant in the spectral fits. This, in combination with the very
low column density of this source and the clear presence of a soft
component in the spectrum \citep[likely due to the accretion disc,
see][]{2009MNRAS.395L..52R}, might have artificially skewed the photon
index to high values. However, since it is unclear what photon index
would have been obtained for the source if the data quality was lower
and the absorption higher, we still include this data point in our
figure. Although in this work we aim to describe the general behaviour
rather than explaining individual sources, it is clear that individual
sources might behave (slightly) differently compared to the majority
of sources.

\section{Discussion \label{discussion}}

We searched the literature for reports on the spectral properties of
neutron-star LMXBs when they have accretion luminosities between
$10^{34}$ and $10^{36}$ ergs s$^{-1}$, corresponding to roughly 0.01\%
to 1\% of the Eddington mass accretion rate ($\dot{M}_{Edd}$). Despite
the variety of models fitted to the X-ray spectra, for many systems
the results were reported from fits that used a simple absorbed
power-law model. When plotting the photon index versus the luminosity
(Fig.~\ref{photonindex}) a clear trend is visible: the photon index
increases (thus the spectra become softer) with decreasing
luminosities. Such behaviour has been reported before for individual
neutron-star LMXBs \citep[e.g., see][and references
therein]{2013ApJ...767L..31D,2011MNRAS.417..659A,2013MNRAS.434.1586A,2013MNRAS.436L..89A},
but here we demonstrate that very likely most neutron-star systems
behave in a similar manner and they might even follow a universal
relation.

When comparing the neutron-star systems with the averaged points
reported by \citet[][]{2013ApJ...773...59P} for a collection of black
hole systems (10 sources) and with the specific black-hole transient
Swift J1357.2--0933 \citep[][]{2013MNRAS.428.3083A}, it is clear that
the black holes are significantly harder at luminosities below
$10^{35}$ erg s$^{-1}$. Not enough black-hole systems are available in
the luminosity range $10^{35}$ to $10^{36}$ erg s$^{-1}$ to make
conclusive statements for this range. Although the black-hole
transients observed in this luminosity range are on average harder
than the neutron-star systems, there is significant overlap between
the two source classes {\it and} the hardest source \citep[IGR
J18245--2452;][]{2014MNRAS.438..251L} in our sample is in fact a
neutron-star LMXB and not a black-hole system.

Similar to the neutron-star systems, at low luminosities the
black-hole binaries also become softer, but this softening occurs at
significantly lower luminosities \citep[around $10^{34}$ erg s$^{-1}$;
see also the discussions
in][]{2013MNRAS.428.3083A,2013ApJ...773...59P} than what we observe
for the neutron-star systems for which the softening already starts at
$10^{36}$ erg s$^{-1}$. In addition, the black holes seem to level off
at a photon index $\sim$2, but the neutron-star systems reach photon
indices of 2.5--3.  Neutron-star systems at even lower luminosities
(below $10^{34}$ erg s$^{-1}$; i.e., when they are in quiescence)
display a large variety of behaviour (see section~\ref{quiescence})
and are therefore not studied in this work.

\subsection{Origin of the softening of the neutron-star systems \label{origin}}

When high quality data is available of the neutron-star systems at
X-ray luminosities in the range $10^{34}$ to $10^{35}$ erg s$^{-1}$,
typically a soft component needs to be added to the power-law model to
obtain an acceptable fit
\citep[e.g.,][]{2013MNRAS.434.1586A,2013ApJ...767L..31D}. It also has
been found that when the X-ray luminosity decreases for certain
sources, the temperature of this soft component decreases. The soft
component and the decrease in temperature have been interpreted
\citep[][]{2013ApJ...767L..31D,2013MNRAS.436L..89A,2014ApJ...780..127B}
as the neutron-star surface becoming clearly visible in the X-ray
spectra originating from low-level accretion onto the neutron-star
surface. The decrease in luminosity is then due to a decrease in
accretion rate onto the surface causing the surface temperature to go
down.

Interestingly, when a soft component is added to the spectral model,
the power-law component suddenly becomes significantly harder with a
photon index well below 2 \citep[e.g.,][we note that the errors on
the photon indices are typically large but the general trend is that
the power-law component is rather
hard]{2012ApJ...759....8D,2013MNRAS.434.1586A,2013ApJ...767L..31D}. This
might suggest that the softening of the overall spectrum is due to the
neutron-star surface becoming more and more dominant in the X-ray
spectra. However, at luminosities above $10^{35}$ erg s$^{-1}$, a soft
component is not always needed to fit the spectra adequately
\citep[][]{2013MNRAS.434.1586A}. In addition, the photon indices are
relatively high already irrespectively of whether or not a soft
component is included in the spectral fits.

\begin{figure*}
 \begin{center}
\includegraphics[width=1.5\columnwidth,angle=-90]{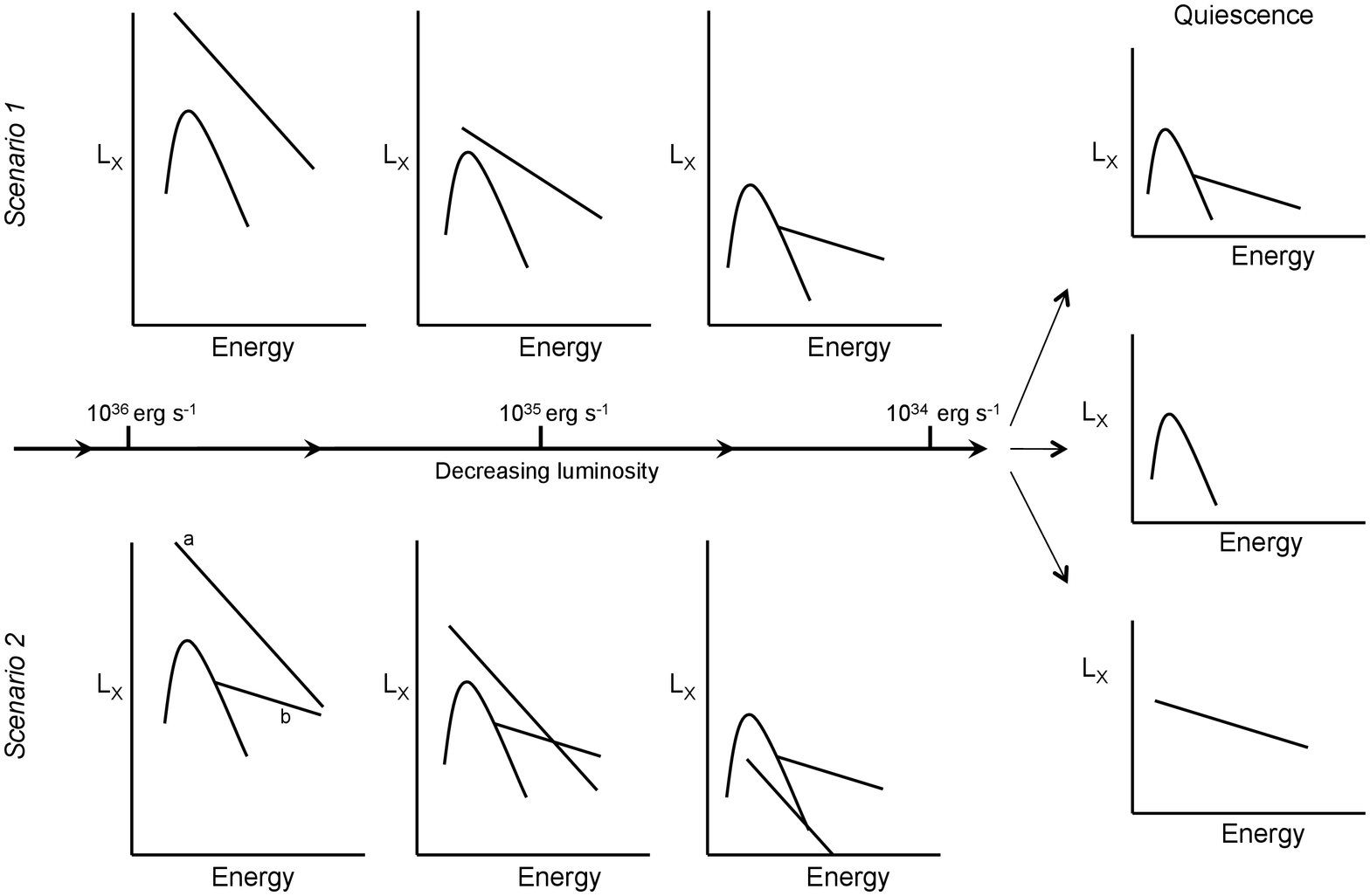}
    \end{center}
 \caption[]{Two possible scenarios explaining the observed softening of the
neutron-star spectra when the X-ray luminosity decreases from
$10^{36}$ erg s$^{-1}$ down to $10^{34}$ erg s$^{-1}$ (see main text
for full details). In quiescence (right panels) three different type of quiescence
spectra are observed: the spectrum is fully dominated by the power-law
component (bottom), the spectrum is fully dominated by the thermal
component (middle) or in the spectrum both components are clearly
detected (top).}
\label{scenarios}
\end{figure*}

Taking the above observational facts into account, we propose two
possible phenomological scenarios \citep[see also the discussion
in][]{2013MNRAS.434.1586A} that might be able to explain the softening
of the neutron-star spectra when the luminosity decreases from
$10^{36}$ erg s$^{-1}$ to $10^{34}$ erg s$^{-1}$. These different
scenarios are illustrated in Figure~\ref{scenarios}. In scenario 1,
the power-law component (assumed to be due to the accretion process)
starts out with a relatively hard index ($<$2; the source is in the
so-called hard state or extreme island state for neutron-star LMXBs)
for luminosities above $10^{36}$ erg s$^{-1}$ (note that this
luminosity range is not shown in Fig~\ref{scenarios}, which starts at
$10^{36}$ erg s$^{-1}$). When the luminosity decreases to
$\sim10^{35}$ erg s$^{-1}$, the photon index becomes larger
(increasing to a value of $>$2; left panel of scenario 1 in
Fig.~\ref{scenarios}) resulting in an overall softening of the
spectrum. The thermal component (assumed to be due to low-level
accretion onto the neutron-star surface) is either still absent or
only weakly detectable in the X-ray spectra and does not contribute to
the softening of the overall spectrum. Then, at luminosities below
$10^{35}$ erg s$^{-1}$ the power-law component becomes suddenly harder
again (with a photon index well below 2 again; two right panels of
scenario 1 in Fig.~\ref{scenarios}). At the same time the thermal
component becomes a major component in the X-ray spectra and since its
temperature is going down with decreasing luminosity, the resulting
overall spectrum softens further.

In scenario 2, the source started out in a similar state as in
scenario 1, with a hard power-law component at a luminosity of
$10^{36}$ erg s$^{-1}$. Also in this scenario the power-law component
(labelled a) softens when the luminosity decreases to $10^{35}$ erg
s$^{-1}$ but around this luminosity the soft component might become
visible in the spectra {\it together} with an extra harder power-law
component (with a photon index well below 2; labelled b in
Fig.~\ref{scenarios}, left panel) whose origin is not known at the
moment.  Due to the usually poor quality of the obtained data, the
components are typically difficult to separate from each other and
when fitting the data with a single power-law model, the overall
spectrum is rather soft with a high photon index ($>2$).  When the
luminosity decreases further, the power-law component (power law a)
due to the accretion flow decreases in strength. Although this
component could possibly soften further, at this time the thermal
emission from the neutron-star surface and the second, harder
power-law component (power law b) dominate the spectrum. At the lowest
luminosities of $\sim10^{34}$ erg s$^{-1}$, the original power-law
component (power law a) has decayed to very low flux levels and cannot
be detected anymore in the spectra. Only the thermal component and the
hard power-law component (power law b) are detected. The softening of
the X-ray spectra is mostly due to a decrease of the temperature of
the thermal component.

Although it is currently unclear which scenario is the correct one (or
that it could be a combination of both scenarios), we currently
consider scenario 2 as the most promising one. In black-hole systems
the power-law component {\it must} come from the accretion process and
it is now well established that at the lowest accretion luminosities
this component becomes softer \citep[see discussion
in][]{2013ApJ...773...59P}. Considering the many similarities in the
accretion process between neutron-star systems and the black-hole
ones, it is reasonable to assume that the power-law component will
behave similarly in both types of systems. The softening can be
explained in context of radiatively inefficient accretion flows
\citep[see the discussions in][and references
therein]{2013MNRAS.428.3083A,2013ApJ...773...59P}. In this model, it
is difficult to understand why in the neutron-star systems the spectra
should be harder than in the black-hole binaries when both systems are
in quiescence. We consider it more plausible that not one but two
processes are active at low accretion rates in the neutron-star
systems which both produce a power-law component in the spectrum. One
process is tied to the accretion flow at large distances and one is
due to an unknown origin (in section~\ref{contribution} we discuss the
possible nature of this component further). High signal-to-noise ratio
observations at the right X-ray luminosity are needed to be able to
distinguish between the two scenarios.

When the luminosity decreases even further (to quiescent luminosities)
for the neutron-star systems, three different possibilities can be
observed (see Fig.~\ref{scenarios}). First, some systems are totally
dominated by a power-law component (Fig.~\ref{scenarios} right
bottom). It is currently still unclear what causes this component
\citep[it has been postulated that it might be related to the magnetic
field of the neutron star; see the discussions
in][]{1998A&ARv...8..279C,2001ApJ...559.1054R,2012ApJ...756..148D} and
if it is related to the power-law component (scenario 1) or components
(scenario 2) seen at higher luminosities. Second, some quiescent X-ray
spectra could be totally dominated by a thermal component
(Fig.~\ref{scenarios} right middle). Typically it is assumed that this
is due to cooling emission from the neutron star which has been heated
during outburst
\citep[e.g.,][]{1998ApJ...504L..95B,1999ApJ...514..945R,2001ApJ...551..921R,2001ApJ...559.1054R,2000A&A...358..583C},
although very low level accretion onto the neutron-star surface cannot
be excluded either. Third, in some quiescent sources both spectral
components are still clearly detectable. Usually, the power-law
component is hard with photon indices well below 2 (and sometimes even
below 1, although the errors are typically very large). Although the
soft component could be due to cooling of the neutron star,
variability in this component in some neutron-star X-ray transients in
quiescence \citep[e.g., Cen X-4, Aql
X-1][]{1997A&A...324..941C,2004ApJ...601..474C,2002ApJ...577..346R,2010ApJ...720.1325C,2011MNRAS.414.3006C,2013MNRAS.433.1362C,2013MNRAS.436.2465B}
suggests that very low-level accretion onto the neutron-star surface
is more likely in those systems. Since the spectral shape (i.e., a two
component model with a hard power-law component) of these quiescent
neutron stars is very similar to that of the neutron-star LMXBs at
$\sim10^{34}$ erg s$^{-1}$, we tentatively suggest that the underlying
physical mechanism is the same. The lower temperature of neutron-star
surfaces in the quiescent systems compared to those at higher
luminosities can be explained because of the lower accretion rate onto
the surface of the neutron stars. The hard component still remains an
enigma, but in the next section we suggest that it might be directly
connected with the accretion process onto the surface as
well. 

This suggestion is further supported by the very recent paper by
\cite{2015MNRAS.449.2803D} which studied the quiescent spectrum of the
neutron-star X-ray transient Cen X-4. They also reached the conclusion
that in this source both the thermal as well as the non-thermal
component are due to low level accretion onto the neutron-star
surface. This conclusion is consistent with our conclusions, although
we suggest that this is true not only for the quiescent systems but
also for the sources that have X-ray luminosities up to a few times
$10^{34}$ erg s$^{-1}$.

\subsection{A roughly equal contribution of the soft and hard component to the 0.5--10 keV luminosity? \label{contribution}}

When both the soft and the hard component are clearly distinguished in
the X-ray spectrum, one can calculate the relative contribution of
both components to the 0.5--10 keV luminosity. This has been done for
several sources, both actively accreting sources
\citep[][]{2013MNRAS.434.1586A,2014MNRAS.441.1984C} and quiescent
systems
\citep[][]{2010ApJ...720.1325C,2011ApJ...736..162F,2014arXiv1408.3276H}. When
looking at those references, it becomes strikingly apparent that when
there are good indications that the source is actively accreting onto
the surface of the neutron star\footnote{This would be because type-I
X-ray bursts are observed or strong variability is observed either
during the observation or between different observations \citep[e.g.,
accretion flares on top of a crust cooling
curve][]{2011ApJ...736..162F,2014arXiv1408.3276H}} in many occasions
both components contribute about half\footnote{With 'about half' we
mean that the contribution of the different components always lay
within 40\% to 60\% of the total flux.}  the flux in the 0.5--10 keV
energy range (within the, often large, errors).

Moreover, for Cen X-4 there are many quiescent observations at
different flux levels and both components increase and decrease in a
similar fashion, ensuring that the contribution of both components to
the 0.5-10 keV flux remains about equal
\citep[][]{2010ApJ...720.1325C}. In addition,
\citet[][]{2014MNRAS.441.1984C} reported for Aql X-1 that the
fractional contribution of both components remained roughly constant
(for luminosities below $10^{35}$ erg s$^{-1}$; above that the
power-law component dominated fully the X-ray spectra) when it was
decaying after a main outburst. Only at the lowest luminosities (a few
times 10$^{33}$ erg s$^{-1}$) the thermal component dominated
(although the power-law contribution was still of order 30\%; at these
luminosities it was proposed by \citet[][]{2014MNRAS.441.1984C} that
the source was in the cooling regime and that the neutron star was not
accreting anymore). Finally, \citet[][]{2014ApJ...780..127B} found
that during the rise of the third discovered X-ray transient in the
globular cluster Terzan 5, the soft component and the hard component
also increased together in such a way that their fractional
contributions to the 0.5--10 keV luminosity remained approximately
50\%. Only above a luminosity of $\sim10^{35}$ erg s$^{-1}$ this broke
down and the power-law component became much more dominant than the
thermal component (see their Table~4).

This all indicates that some fundamental physical process is occurring
at low accretion rates in neutron-star systems that causes the
physical mechanism behind the hard component to be connected with that
behind the soft component. Only above a luminosity of $\sim10^{35}$
erg s$^{-1}$ this might not be true anymore since the power-law
component then dominates the X-ray spectra. At quiescent luminosities
($<10^{34}$ erg s$^{-1}$) it depends on the contribution of the
thermal component due to the cooling of the neutron star, which
physical process dominates the X-ray emission. It is interesting to
note that in the most basic disc theory for accretion onto neutron
stars \citep[][]{2002apa..book.....F}, half the luminosity is
generated in the accretion disc and the other half when the matter
hits the surface. Although it is likely that no disc is present close
to the neutron stars when they are at very low luminosities, it is
possible that the energy stored in the accretion flow is released in
such a way that half of it is emitted very close to the neutron star
and the remainder when the matter hits the star.

Only a few authors have modelled low level accretion onto neutron
stars \citep[see for example the discussions
in][]{2002ApJ...580..413R,2014ApJ...780..127B}. \citet[][]{1995ApJ...439..849Z}
found that the resulting X-ray spectrum would be harder than a simple
black-body spectrum, but no significant hard tail would be
present. Similar conclusions were reached by other authors although
some studies found that a (relatively weak) power-law component was
also present \citep[e.g.,][]{2001A&A...377..955D}. More detailed
theoretical investigations are needed to determine the exact emerging
spectra of low-level accretion onto neutron stars, but if our
hypothesis is correct the hard power-law component also have to be
explained by such models and that it should contribute about half to
the 0.5--10 keV luminosity.

\subsection{Comparing with  Jonker et al. (2004a,b)}

\citet[][]{2004MNRAS.349...94J,2004MNRAS.354..666J} found that for
quiescent neutron-star systems there appears to be an anticorrelation
between the fractional power-law contribution and the total 0.5--10
keV luminosity for systems that are fainter than $\sim2 \times
10^{33}$ erg s$^{-1}$ and a correlation between those contributions
for systems that are brighter than this luminosity. From Figure 5 in
\citet[][]{2004MNRAS.354..666J} it can be seen that they have data
points for only one source that has luminosities above $10^{34}$ erg
s$^{-1}$ (i.e., XTE J1706--267). Consistent with what we discussed in
section \ref{contribution}, that source has a fractional contribution
of the power-law component of $\sim$50\%. The reason why, when the
luminosity decreases below $10^{34}$ erg s$^{-1}$, the fractional
contribution of the power-law component decreases is likely that the
thermal emission will be dominated by cooling emission from the
neutron-star surface and not by the accretion emission. If true, then
the exact relation might be different for different sources since the
temperature of the neutron-star surface is likely not the same for all
sources because of their different accretion histories. The emergence
of a dominating power-law component below $2\times 10^{33}$ erg
s$^{-1}$ might suggest that at the lowest luminosity a different
power-law component appears (as suggested as one of the possibilities
in section \ref{contribution}), whose origin is still unclear.

\subsection{A diagnostic tool to separate the neutron-star systems
from the black-hole ones? \label{NS_candidates}}

The strong indication that most neutron-star LMXBs are significantly
softer than black-hole systems below an X-ray luminosity of $10^{35}$
erg s$^{-1}$ (and possibly $10^{36}$ erg s$^{-1}$, but at this
luminosity the source classes overlap) suggests that we can use the
rough spectral shape between 0.5 and 10 keV as a diagnostic tool to
separate the neutron-star LMXBs from the black-hole systems. For
example, for both IGR J17494-3030 and XTE J1719-291 it has not
conclusively been determined that they are neutron-star systems since
no type-I X-ray bursts or X-ray pulsations were observed. However, it
has been argued that they are strong candidate neutron-star LMXBs
\citep{2011MNRAS.417..659A,2013MNRAS.436L..89A}. Indeed, the spectral
index obtained when fiting their spectra with a simple power-law model
falls right in the regime of the confirmed neutron-star
LMXBs\footnote{We note that if we remove both sources from our plots,
the correlation we found does not change so the effect of those
sources on our correlation is minimal.}.

Similarly, \citet[][]{2011MNRAS.415.2373S} reported on the X-ray
spectral properties of the unclassified X-ray transient IGR
J17285--2922 (also called XTE J1728--295). From the low X-ray upper
limit on the quiescent luminosity of the source, they suggested it
likely harbours a black hole as compact primary, but they could not be
conclusive. However, they also reported that the photon index was
1.61$\pm$0.01 when the source had a X-ray luminosity of $\sim6.1\times
10^{35}$ erg s$^{-1}$ (0.5--10 keV; converted from the listed 0.3--10
keV flux using WebPIMSS, and assuming a distance of 8 kpc). This would
place the source just below the neutron-star points in
Figure~\ref{photonindex} but fully consistent with the black-hole
points. Although the neutron-star transient IGR J18245--2452 in the
globular cluster M28 has similar hard spectra at this luminosity
\citep[][]{2014MNRAS.438..251L}, this behaviour is more commonly
observed for black-hole systems.  So, although we cannot conclusively
determine the nature of the compact object in this system, this would
add evidence to the suggestion by \citet[][]{2011MNRAS.415.2373S} that
this source harbours a black hole.

Another interesting source is AX
J1548.3--5541. \citet[][]{2012AA...540A..22D} suggested that it is
likely a LMXB. Its X-ray spectrum is too soft (with a photon index of
2.3$\pm$0.1) for it to be a high-mass X-ray binary. These authors
could not say anything about the nature of the accretor if the source
is truly a LMXB. In their paper, they reported the luminosity in the
0.3--10 keV band and when we convert that (using WebPIMMS) to 0.5--10
keV we get a luminosity of $\sim3 \times 10^{35}$ erg s$^{-1}$. These
luminosity and photon index would place the source right on the
neutron-star track in Figure~\ref{photonindex} and therefore, we
tentatively classify this source as a neutron-star LMXB.

Another unclassified source is the X-ray transient XMMSL1
J171900.4-353217
\citep[see][]{2010ATel.2607....1R,2010ATel.2627....1R,2010ATel.2615....1M}.
\citet[][]{2010ATel.2722....1A} and \citet[][]{2010ATel.2627....1R}
reported on the spectral properties of this source and they found a
relatively soft X-ray spectrum (with photon index $>$2). If we assume
a fiducial distance of 8 kpc for this source, its X-ray luminosity is
typically around $10^{35}$ erg s$^{-1}$
\citep[][]{2010ATel.2627....1R,2010ATel.2722....1A}.  This would put
the source right on the neutron-star track and well above that
observed for the black-hole systems. Therefore, we also tentatively
identify this source as a neutron-star LMXB.

A very peculiar source is the X-ray transient IGR J17361--4441
located in the globular cluster NGC 6388
\citep[][]{2011ATel.3565....1G,2011ATel.3566....1F}. The nature of
this source remains an enigma. Its transient behaviour, and location
in a globular cluster would suggest the source being a LMXB. However,
its very hard X-ray spectra \citep[with photon indices
$\sim$1;][]{2011ATel.3566....1F,2011ATel.3595....1W,2011A&A...535L...1B}
are not consistent with the typically observed spectra of LMXBs at the
luminosities observed for the source \citep[peak luminosities of 6 to
9 $\times10^{35}$ erg s$^{-1}$;][]{2011ATel.3595....1W}. Alternative
explanations have been put forward
\citep[][]{2011ATel.3595....1W,2014MNRAS.444...93D}, but in the
appendix we argue that the source is indeed a LMXB harbouring either a
black hole or a low-magnetic field neutron star. However, comparing
IGR J17361--4441 with the other sources shown in
Figure~\ref{photonindex} demonstrates that it is not consistent with
either the neutron-star systems or with the black-hole
transients. Therefore, some unusual systems behave differently than
the average population and are not classifiable using
Figure~\ref{photonindex}.

To conclude, we propose that comparing the spectral index obtained
from fits with a simple power-law model at a 0.5--10 keV luminosity of
$10^{34}-10^{35}$ erg s$^{-1}$ (Fig.~\ref{photonindex}) can be used to
suggest in most circumstances (but not all, as we demonstrated above
using IGR J173611--4441) the nature of the compact object in an
unclassified X-ray binary, if this source exhibits accretion
luminosities between $10^{34}$ and $10^{35}$ erg s$^{-1}$. More
systems need to be studied in this luminosity range to confirm that
Figure~\ref{photonindex} can indeed be used as a diagnostic tool. In
addtion, more systems have to be studied between $10^{35}$ and
$10^{36}$ erg s$^{-1}$ to determine if also in this luminosity range
neutron-star LMXBs have (on average) softer spectra than the
black-hole systems.

\vspace{1cm}

\noindent {\bf Acknowledgements}\\

RW and MAP acknowledge support from a European Research Council (ERC)
starting grant awarded to RW. ND acknowledges support via an EU Marie
Curie Intra-European fellowship under contract
no. FP-PEOPLE-2013-IEF-627148. MAP is supported by Canary Island CIE:
Tricontinental Atlantic Campus. DA acknowledges support from the Royal
Society. COH is suppored by an NSERC Discovery Grant and an Ingenuity
New Faculty Award and an Alexander von Humboldt Fellowship. RW thanks
Caroline D'Angelo for useful discussion and for receiving her
manuscript before submission. This research has made use of NASA's
Astrophysics Data System.

\bibliographystyle{mn2e}

\newpage\clearpage
\begin{table}
\caption{Neutron-star X-ray transients which have low N$_{\rm H}$ \label{sources}}
\begin{tabular}{llcccc}
\hline
Source                 & References                      & Distance      & N$_{\rm H}^{h}$ & $\Gamma$ & L$_{\rm X} $ \\
                       &                                 &  (kpc)        &   ($10^{22}$ cm$^{-2}$) & & ($10^{35}$ erg s$^{-1}$) \\
\hline
AX J1754.2--2754       & \citet[][]{2013MNRAS.434.1586A} & 9.2		 & 2.93$\pm$0.06  & 2.51$\pm$0.03 & 2.74$\pm$0.08 	 \\ 
                       & \citet[][]{2012AA...540A..22D}$^a$  &               & 2.1$\pm$0.3    & 2.8$\pm$0.2   & 2.1$\pm$0.2    \\
1RXS J171824.2--402934 & \citet[][]{2013MNRAS.434.1586A} & 9.0           & 1.78$\pm$0.04 & 2.33$\pm$0.04 & 0.78$\pm$0.01  \\ 
                       & \citet[][]{2009AA...506..857I}  &               & 1.3 & 2.44$\pm$0.1  & 0.63$\pm$0.03 \\
                       &                                 &               &     & 2.37$\pm$0.11 & 0.54$\pm$0.03  \\
                       & \citet[][]{2009ApJ...699.1144C}$^{b,c}$ &               & 1.2$\pm$0.1 & 2.3$\pm$0.3 & 0.88  \\
                       &                                 &               &             & 1.9$\pm$0.1 & 3.6  \\
                       &                                 &               &             & 2.1$\pm$0.1 & 2.1  \\
                       &                                 &               &             & 2.4$\pm$0.1 & 0.97  \\
                       &                                 &               &             & 2.4$\pm$0.2 & 0.56  \\
1RXH J173523.7--354013 & \citet[][]{2013MNRAS.434.1586A} & 9.5           & 1.43$\pm$0.07 & 2.45$\pm$0.07 & 0.56$\pm$0.03  \\ 
XTE J1709--267         & \citet[][]{2013ApJ...767L..31D} & 8.5           &0.34$\pm$0.03  &2.0$\pm$0.1 & 29$\pm$2 	 \\ 
                       &                                 &               &               &2.4$\pm$0.1 & 15$\pm$2 \\
                       &                                 &               &               &2.6$\pm$0.3 & 0.7$\pm$0.1 \\
                       &                                 &               & 0.49$\pm$0.01 &2.90$\pm$0.04 & 0.201$\pm$0.001$^e$\\
IGR J17062-6143        & \citet[][]{2012ATel.4219....1D}$^c$ & 5.0		& 0.20$\pm$0.01  &2.1$\pm$0.1  & 6	 \\ 
1RXS J170854.4--321857 & \citet[][]{2005AA...440..287I}   & 13.0		 &0.40$\pm$0.01 & 1.9$\pm$0.2  & 15$\pm$1 	\\ 
                       & \citet[][]{2009ApJ...699.1144C}$^c$ &                       &0.43$\pm$0.02 & 1.98$\pm$0.05 & 33\\
                       &                                 &                       &              & 2.4$\pm$0.1   &3.6\\
SAX J1753.5-2349       & \citet[][]{2009ApJ...699.1144C}$^c$ & 8.8               & 1.9$^{+0.5}_{-0.6}$ & 2.0$\pm$0.4 & 7.7 \\
Swift J174805.3-24463$^f$ & \citet[][]{2014ApJ...780..127B}$^g$ & 5.9         & 1.74  & 2.5$\pm$0.4 & 1.0$^{+0.3}_{-0.2}$  \\
                       &                                 &                  &    & 2.3$\pm$0.3 & 1.5$\pm$0.2\\
                       &                                 &                  &    & 1.9$\pm$0.4 & 3.7$\pm$0.5\\
                       &                                 &                  &    & 1.4$\pm$0.1 & 27.3$\pm$1.3\\
                       &                                 &                  &  & 2.07$\pm$0.07 & 21.3$\pm$0.8\\
                       &                                 &                  &    & 1.7$\pm$0.1 & 12.2$\pm$0.7\\
                       &                                 &                  &    & 1.8$\pm$0.1 & 10.8$\pm$0.7\\
                       &                                 &                  &    & 1.9$\pm$0.2 & 11.0$\pm$1.2\\
                       &                                 &                  &    & 1.8$\pm$0.1 & 9.1$\pm$0.5\\
                       &                                 &                  &    & 1.9$\pm$0.2 & 8.9$\pm$0.7\\
                       &                                 &                  &    & 2.0$\pm$0.1 & 6.2$\pm$0.5\\
                       &                                 &                  &    & 1.8$\pm$0.3 & 4.6$\pm$0.5\\
                       &                                 &                  &    & 2.0$\pm$0.2 & 1.9$\pm$0.2\\
Aql X-1                & \citet[][]{gandhi2014}          & 5.0                 & 0.95$\pm$0.2 & 2.6$\pm$0.3 & 0.60$\pm$0.15\\
\hline
IGR J17494--3030$^d$   & \citet[][]{2013MNRAS.436L..89A} & 8.0           & 1.87& 1.8$\pm$0.2 & 10.7$\pm$0.08 	 \\ 
                       &                                 &               &     & 1.9$\pm$0.1 & 13.4$\pm$0.08 \\
                       &                                 &               &     & 1.8$\pm$0.3 & 8.7$\pm$0.08 \\
                       &                                 &               &     & 2.0$\pm$0.3 & 2.6$\pm$0.08 \\
XTE J1719--291$^d$     & \citet[][]{2011MNRAS.417..659A} & 8.0           & 0.53 & 2.02$\pm$0.08 & 13.3$\pm$0.8          \\ 
                       &                                 &               &      & 2.74$\pm$0.05 & 0.475$\pm$0.007 \\
                       &                                 &               &      & 2.83$\pm$0.25 & 0.36$^{+0.03}_{-0.02}$ \\
                       &                                 &               &      & 2.6$\pm$0.4 & 0.30$^{+0.06}_{-0.03}$ \\
                       &                                 &               &      & 2.32$\pm$0.11 & 2.8$\pm$0.2 \\
                       &                                 &               &      & 2.15$\pm$0.09 & 4.4$\pm$0.3 \\
                       &                                 &               &      & 2.7$\pm$0.4 & 0.38$^{+0.07}_{-0.03}$ \\
\hline
\multicolumn{6}{l}{$^a$ 0.3--10 keV fluxes converted to 0.5--10 keV fluxes using WebPIMMS}\\
\multicolumn{6}{l}{$^b$ Luminosities converted to 9 kpc}\\
\multicolumn{6}{l}{$^c$ No errors on luminosities (see section~\ref{errorsLx})}\\
\multicolumn{6}{l}{$^d$ NS nature not yet confirmed}\\
\multicolumn{6}{l}{$^e$ Luminosity calculated from a power-law plus a neutron-star atmosphere model fit to the same spectrum} \\
\multicolumn{6}{l}{$^f$ Also know as Terzan 5 X-3; the third bright X-ray transient in the globular cluster Terzan 5}\\
\multicolumn{6}{l}{$^g$ Data taken from their Table 9; excluding the data with X-ray luminosities $>5\times 10^{36}$ erg s$^{-1}$ and with errors on the photon index $>$0.5}\\
\multicolumn{6}{l}{$^h$ The column density as obtained in the quoted papers. If no errors are given, the column density was fixed to the value quoted (i.e., for }\\
\multicolumn{6}{l}{{1RXS J171824.2--402934 the $N_{\rm H}$ used was obtained by \citet{2005AA...440..287I} using a higher quality Chandra observation of the source; }}\\
\multicolumn{6}{l}{{for Swift J174805.3--24463 the $N_{\rm H}$ used was obtained from quiescent spectra also reported by \citet{2014ApJ...780..127B}; for IGR J17494--3030 and}}\\
\multicolumn{6}{l}{{XTE J1719--291 the reported $N_{\rm H}$ was deteremined from high quality XMM-Newton data reported in \citet{2013MNRAS.436L..89A} and}}\\
\multicolumn{6}{l}{{\citet{2011MNRAS.417..659A}, respectively).}}\\
\end{tabular}
\label{tab1}
\end{table}

\newpage\clearpage
\begin{table}
\caption{Black-hole X-ray transients \label{sourcesBH}}
\begin{tabular}{llcccc}
\hline
Source                 & References                      & Distance      & N$_{\rm H}^{f}$ & $\Gamma$ & L$_{\rm X} $ \\
                       &                                 &  (kpc)        &   ($10^{22}$ cm$^{-2}$) & & ($10^{35}$ erg s$^{-1}$) \\
\hline
Swift J1357.2--0933$^{a,b}$ & \citet[][]{2013MNRAS.428.3083A}&1.5		&0.012 & 1.53$\pm$0.02 & 1.13$\pm$0.02 	\\ 
                             &                                &                 &      & 1.58$\pm$0.02 & 1.01$\pm$0.01 \\
                             &                                &                 &      & 1.56$\pm$0.01 & 0.941$\pm$0.007 \\
                             &                                &                 &      & 1.55$\pm$0.02 & 0.708$\pm$0.008 \\
                             &                                &                 &      & 1.57$^{+0.05}_{-0.06}$ & 0.56$\pm$0.02 \\
                             &                                &                 &      & 1.59$\pm$0.02 & 0.445$\pm$0.005 \\
                             &                                &                 &      & 1.58$\pm$0.02 & 0.346$\pm$0.004 \\
                             &                                &                 &      & 1.61$\pm$0.03 & 0.203$\pm$0.004 \\
                             &                                &                 &      & 1.65$\pm$0.05 & 0.110$\pm$0.003 \\
                             &                                &                 &      & 1.78$\pm$0.06 & 0.059$\pm$0.002 \\
                             &                                &                 &      & 1.9$\pm$0.2 & 0.012$\pm$0.001 \\
Plotkin black-hole sample$^{c,d}$ & \citet[][]{2013ApJ...773...59P} & --  	      	  & --& 2.05$^{+0.21}_{-0.17}$ & $(4.1\pm2.5) \times 10^{-5}$  	\\ 
                             &                                           &                &   & 2.15$^{+0.19}_{-0.18}$ & $(3.3\pm1.8) \times 10^{-4}$\\
                             &                                           &                &   & 1.98$^{+0.10}_{-0.11}$ & $(4.4\pm1.7) \times 10^{-3}$\\
                             &                                           &                &   & 2.02$\pm$0.06          & $(3.7\pm2.1) \times 10^{-2}$\\
                             &                                           &                &   & 1.67$\pm$0.08          & $0.43\pm0.09$\\
\hline
Swift J1753.5--0217          & \citet[][]{2010MNRAS.402..836R}$^e$ &  8.5           & 0.175$\pm$0.001 & 1.666$\pm$0.003  & 35.44$\pm$0.09   \\
GRO J1655--40                & \citet[][]{2010MNRAS.402..836R}$^e$ &  3.2           & 0.525$\pm$0.003 & 1.660$\pm$0.005  & 11.6$\pm$0.2   \\
XTE J1118+480                & \citet[][]{2010MNRAS.402..836R}$^e$ &  1.72          & 0.0080$\pm$0.0001& 1.877$\pm$0.005 & 4.60$\pm$0.04   \\
MAXI J1659--152              & \citet[][]{2012MNRAS.423.3308J} &  6             & 0.27$\pm$0.01 & 1.48$\pm$0.03 & 0.82$\pm$0.09\\
H 1743--322                  & \citet[][]{2010MNRAS.401.1255J} &  8.5           & 2.3  & 1.74$\pm$0.05 & 25.3$\pm$0.5\\
                             &                                 &                &      & 1.59$\pm$0.08 & 9.2$\pm$0.3\\            
                             &                                 &                &      & 1.70$\pm$0.08 & 3.29$\pm$0.09\\            
                             &                                 &                &      & 1.6$\pm$0.1 & 1.82$\pm$0.09\\            
                             &                                 &                &      & 1.9$\pm$0.2 & 0.66$\pm$0.05\\            
\hline
\multicolumn{6}{l}{$^a$ Strong BH candidate}\\
\multicolumn{6}{l}{$^b$ Using the averaged data; data points not tabulated in \citet[][]{2013MNRAS.428.3083A}}\\
\multicolumn{6}{l}{$^c$ Using the averaged data as caculated from their Table 3 and 5}\\
\multicolumn{6}{l}{$^d$ The luminosity errors correspond to the standard deviation on the luminosity points use to calcualte the average}\\
\multicolumn{6}{l}{$^e$ 0.5-10 keV fluxes obtained using a more complex model.}\\
\multicolumn{6}{l}{$^f$ {The column density as obtained in the quoted papers. If no errors are given, the column density was fixed to the value quoted (i.e., for }}\\
\multicolumn{6}{l}{ {Swift J1357.2--0933 the $N_{\rm H}$ used was obtained by \citet{2014MNRAS.439.3908A} using a high quality XMM-Newton observation of the source; }}\\
\multicolumn{6}{l}{{for H 1743--322 the $N_{\rm H}$ used was obtained from high quality outburst spectra reported by \citet{2006ApJ...646..394M}.)}}\\
\end{tabular}
\label{tabBH}
\end{table}

\newpage\clearpage
\begin{table}
\caption{Neutron-star X-ray transients which are also AMXPs \label{sourcesAMXP}}
\begin{tabular}{llcccc}
\hline
Source                 & References                      & Distance      & N$_{\rm H}^{e}$ & $\Gamma$ & L$_{\rm X} $ \\
                       &                                 &  (kpc)        &   ($10^{22}$ cm$^{-2}$) & & ($10^{35}$ erg s$^{-1}$) \\
\hline
NGC 6440 X-2           & \citet[][]{2010ApJ...714..894H}$^a$ & 8.5       & 0.59                & 2.4$\pm$0.5   & 0.33$\pm$0.08 \\
                       &                                     &           &                     & 2.3$\pm$0.4   & 0.31$\pm$0.06\\
                       &                                     &           &                     & 1.8$\pm$0.2   & 5.9$\pm$0.6\\
                       &                                     &           &                     & 1.9$\pm$0.3   & 1.0$^{+0.2}_{-0.1}$\\
                       &                                     &           &                     & 1.7$\pm$0.1   & 5.5$\pm$0.3\\
                       &                                     &           &                     & 2.3$^{+0.3}_{-0.4}$   & 0.9$\pm$0.1\\
                       &                                     &           & 0.69$\pm$0.06       & 1.8$\pm$0.1   & 15.5$\pm$0.6\\
IGR J00291+5934        & \citet[][]{2010AA...517A..72L}$^b$    &  3        & 0.6$\pm$0.1         & 1.7$\pm$0.1   & 5.2$\pm$0.2\\
                       &                                     &           & 0.5$\pm$0.1         & 1.6$\pm$0.1   & 2.1$\pm$0.2\\
IGR J18245--2452       & \citet[][]{2014MNRAS.438..251L}$^{c}$ &  5.5          & 0.32$\pm$0.02   & 1.34$\pm$0.04 & 18.7$\pm$0.9 \\
                       &                                     &               & 0.44$\pm$0.02   & 1.37$\pm$0.02 & 36.9$\pm$1.1 \\
                       &                                     &               & 0.33$\pm$0.02   & 1.31$\pm$0.04 & 40.2$\pm$2.0 \\
                       &                                     &               & 0.38$\pm$0.02   & 1.26$\pm$0.03 & 48.1$\pm$2.2 \\
                       &                                     &               & 0.38$\pm$0.04   & 1.23$\pm$0.05 & 28.7$\pm$2.1 \\
                       &                                     &               & 0.43$\pm$0.04   & 1.47$\pm$0.05 & 15.5$\pm$1.1 \\
                       &                                     &               & 0.30$\pm$0.08   & 1.50$\pm$0.14 & 1.9$\pm$0.3 \\
                       &                                     &               & 0.51$\pm$0.06   & 1.30$\pm$0.08 & 7.6$\pm$0.8 \\
                       &                                     &               & 0.46$\pm$0.02   & 1.50$\pm$0.03 & 32.8$\pm$1.1 \\
                       &                                     &               & 0.42$\pm$0.04   & 1.53$\pm$0.07 & 6.9$\pm$0.6 \\
                       &                                     &               & 0.37$\pm$0.04   & 1.41$\pm$0.07 & 12.7$\pm$1.1 \\
                       &                                     &               & 0.51$\pm$0.07   & 1.66$\pm$0.10 & 7.8$\pm$0.9 \\
                       &                                     &               & 0.41$\pm$0.02   & 1.46$\pm$0.03 & 17.5$\pm$0.8 \\
                       &                                     &               & 0.39$\pm$0.02   & 1.37$\pm$0.02 & 27.0$\pm$0.9 \\
                       &                                     &               & 0.34$\pm$0.02   & 1.51$\pm$0.03 & 12.2$\pm$0.5 \\
                       &                                     &               & 0.44$\pm$0.04   & 1.56$\pm$0.05 & 5.4$\pm$0.4 \\
                       &                                     &               & 0.4$\pm$0.1$^d$     & 2.4$\pm$0.2 & 0.4$\pm$0.2 \\
\hline
\multicolumn{6}{l}{$^a$ Data points taken from their Table 1 using the same criteria as for the non-pulsating sources.}\\
\multicolumn{6}{l}{$^b$ Data points taken from their Table 6; 2--10 keV fluxes  converted to 0.5--10 keV fluxes using WebPIMMS.}\\
\multicolumn{6}{l}{$^c$ Data points not tabulated by \citet[][]{2014MNRAS.438..251L}.}\\
\multicolumn{6}{l}{$^d$ We used the average data for the observations taken between  April 15-17, 2013.}\\
\multicolumn{6}{l}{$^e$ {The column density as obtained in the quoted papers. If no errors are given, the column density was fixed to the value quoted (i.e., for }}\\
\multicolumn{6}{l}{{NGC 6440 X-2 the $N_{\rm H}$ used was fixed to the cluster value.) }}\\
\end{tabular}
\label{tabAMXPs}
\end{table}

\newpage\clearpage
\begin{table}
\caption{Neutron-star X-ray transients which have high N$_{\rm H}$ \label{sourcesNShighNH}}
\begin{tabular}{llcccc}
\hline
Source                 & References                      & Distance      & N$_{\rm H}^{c}$ & $\Gamma$ & L$_{\rm X} $ \\
                       &                                 &  (kpc)        &   ($10^{22}$ cm$^{-2}$) & & ($10^{35}$ erg s$^{-1}$) \\
\hline
GRS 1741--2853         & \citet[][]{2012AA...545A..49D}$^a$  &  7.2          & 11.4$\pm$1.1    & 2.0$\pm$0.3 & 2.8$\pm$0.3 \\
                       &                                     &                &                 & 1.5$\pm$0.2 &3.3$\pm$0.2\\
                       &                                     &                 &                & 1.6$\pm$0.4 & 0.76$\pm$0.04\\
AX J1745.6--2901       & \citet[][]{2012AA...545A..49D}$^a$  &  8            & 21.8$\pm$0.3     & 1.9$\pm$0.1 & 8.0$\pm$0.3\\
                       &                                     &               &                  & 1.6$\pm$0.1 & 39.8$\pm$0.9\\
                       &                                     &               &                  & 1.8$\pm$0.1 & 32.4$\pm$0.9\\
                       &                                     &               &                  & 1.6$\pm$0.1 & 42.2$\pm$0.9\\
                       &                                     &               &                  & 1.8$\pm$0.1 & 39.0$\pm$1.0\\
                       &                                     &               &                  & 1.6$\pm$0.1 & 17.3$\pm$0.5\\
SAX J1747.0--2853      & \citet[][]{2012AA...545A..49D}$^a$  & 6.7           &  9.5$\pm$0.2     & 2.0$\pm$0.1 &  39.1$\pm$0.5\\
                       &                                     &               &                  & 2.6$\pm$0.1 & 18.6$\pm$0.2\\
KS 1741--293           & \citet[][]{2012AA...545A..49D}$^a$  &  8            &  16.6$\pm$1.8    & 1.8$\pm$0.3 & 2.2$\pm$0.2\\
XMMU J1747161--281048  & \citet[][]{2007AA...468L..17D}$^a$  &  8.4          &  8.9$\pm$0.5      & 2.1$\pm$0.1 & 1.15$\pm$0.07\\
                       & \citet[][]{2011MNRAS.414L.104D}$^{a,b}$ & 8.4           & 8.6$\pm$2.3      & 2.2$\pm$0.5 & 1.2\\
\hline
\multicolumn{6}{l}{$^a$ 2--10 keV fluxes converted to 0.5--10 keV fluxes using WebPIMMS}\\
\multicolumn{6}{l}{$^b$ No errors on the flux (see section~\ref{errorsLx})}\\
\multicolumn{6}{l}{$^c$ The column density as obtained in the quoted papers}\\
\end{tabular}
\label{tabNh}
\end{table}

\newpage\clearpage

\appendix

\section{The transient in NGC 6388 \label{ngc6388}}

The transient IGR J17361--4441 was discovered using INTEGRAL in August
2011 \citep[][]{2011ATel.3565....1G}. The source position is
consistent with that of the globular cluster NGC 6388
\citep[][]{2011ATel.3566....1F}. Its location in a globular cluster
suggests that if the source is an X-ray binary, the accretor is
accreting matter from a low-mass companion star. {\it Swift}/XRT
follow-up observations showed a very hard source with a photon index
$\sim$1
\citep[][]{2011ATel.3566....1F,2011ATel.3595....1W,2011A&A...535L...1B}. Such
a hard index is atypical for neutron-star or black-hole X-ray
transients. It has been suggested it could be a X-ray transient
harbouring a strong magnetic field neutron star
\citep[][]{2011ATel.3595....1W} since such systems have typically
similarly hard spectra
\citep[e.g.,][]{1977ApJ...214..879B,1996ApJ...469L..25G}. Alternatively
it could be an intermediate mass black hole
\citep[][]{2011ATel.3595....1W} but the source position is not exactly
in the center of the cluster and therefore this scenario is unlikely
\citep[][]{2011ATel.3627....1P}. Recently, it was suggested
\citep[][]{2014MNRAS.444...93D} that the transient could be a tidal
disruption event, in which a planetary sized body was disrupted by an
heavy white dwarf (with a mass close to the Chandrasekhar limit).

At the time when the transient was discovered with INTEGRAL, we
requested a DDT observation on {\it XMM-Newton} to study the X-ray
spectrum of this source in detail. This request was approved and the
observation was performed on 23 September 2011 for an on-source
exposure time of $\sim$ 43 ksec. The spectral results of this
observation will be reported elsewhere (Armas Padilla et al.~2015 in
prepration) but also during the {\it XMM-Newton} observation the
source displayed a very hard spectrum (consistent with the Swift/XRT
results). During the {\it XMM-Newton} observation we used the EPIC-pn
camera in timing mode to avoid pile-up and to be able to search for
X-ray pulsations and aperiodic rapid X-ray variability.  We find no
evidence for coherent pulsations, however, we do detect strong
aperiodic variability (including a quasi-periodic oscillation or QPO).

\begin{figure}
 \begin{center}
\includegraphics[width=0.5\columnwidth,angle=-90]{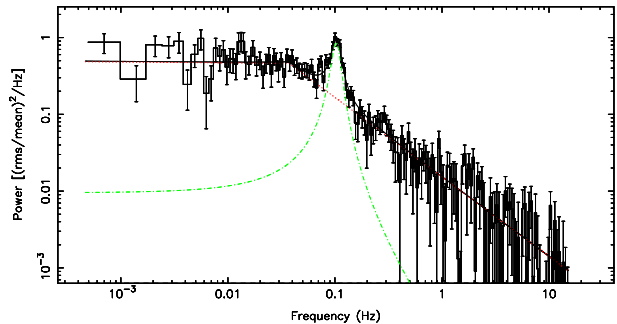}
    \end{center}
 \caption[]{The power density spectrum obtained using the {\it XMM-Newton} pn timing data of the transient in NGC 6388.}
 \label{QPO}
\end{figure}

We applied the standard reduction on the pn data. Since the pn was
used in timing mode, one of the spatial dimensions was collapsed and
the normal practice of extracting the source data using a circle
around the source position cannot be applied. Therefore, as source and
backround events we extracted the data using the RAWX columns [36:39]
and [9:12], respectively. Both the source and the background were
extracted in the energy range 0.5--10 keV. We rebinned the pn data to
a time-resolution of 1 msec and then used a fast fourier transform to
create $\sim$4300 s long power density spectra of the source. All data
were added to create one spectrum. The spectrum was renormalized using
the background count rate obtained from the background data. The
Possion level was estimated from frequencies above 15 Hz and then
subtracted from the power spectrum. The resulting power spectrum is
shown in Figure ~\ref{QPO}. Clearly strong band limited noise is seen
together with a narrow QPO at around 0.01 Hz. We fit the QPO with a
Lorentzian and the band-limited noise with a broken power-law
model. The errors were estimated using $\Delta \chi^2 = 1.0$. For the
QPO we obtained a frequency of 0.1026$\pm$0.0009 Hz, a fractional rms
amplitude of 17.2\%$\pm$0.7\% (0.5--10 keV), and a FWHM of
0.022$\pm$0.002 Hz. For the band-limited noise we obtained a break
frequency of 0.037$\pm$0.004 Hz, an index of 0.01$\pm$0.06 below the
break and 1.03$\pm$0.04 above the break and a fractional rms amplitude
(integrated over 0.001 to 100 Hz; 0.5--10 keV) of 37\%$\pm$2 \%.

\begin{figure}
 \begin{center}
\includegraphics[width=0.5\columnwidth,angle=-90]{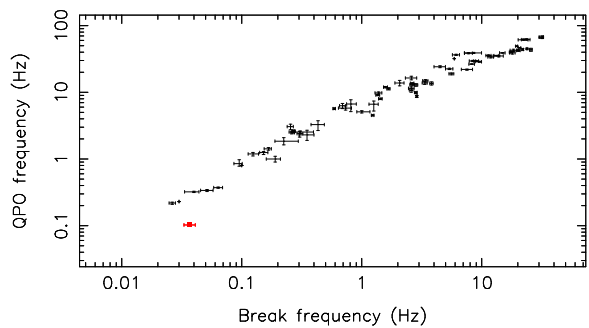}
    \end{center}
 \caption[]{The break frequency versus the QPO frequency. The black points are from \citet[][]{1999ApJ...514..939W} and the red point is for the transient in NGC 6388. (A colour version of this figure is available in the online version of this paper.)}
 \label{WK}
\end{figure}

The power spectrum of the source is a typical one seen often from
low-mass X-ray binaries (both neutron-star as well as black-hole
systems) accreting at relatively low luminosities. The indices of the
band-limited noise are very typical: $\sim$0 before the break
frequency and $\sim$1 above. The strength of the noise is also very
typical. The break frequency and the QPO frequency are very low, but
also what has been seen before in X-ray binaries. To highlight this,
we plotted the QPO frequency versus the break frequency in Figure
\ref{WK} and compared it with the data points reported by
\citet[][]{1999ApJ...514..939W}. Although the QPO frequency seems a
bit low compared to the other sources, the source is still consistent
with the Wijnands \& van der Klis relation when taking into account
the scatter \citep[see the discussion in][]{1999ApJ...514..939W}.

In accreting high-magnetic field neutron-star systems, often
low-frequency QPOs are also observed
\citep[e.g.,][]{1990A&A...230..103B,1990PASJ...42L..27S,1998ApJ...492..342C,2008ApJ...685.1109R,2010MNRAS.407..285J,2011MNRAS.410.1489J}. Although
also band-limited noise is often observed, it does not have the
typical broken power law shape we observe for IGR
J17361--4441. Therefore, we think it is unlikely that the source is a
high-magnetic field neutron-star system. The aperiodic variability of
the source shows that most likely the source is an accreting neutron
star or black hole in a low-mass X-ray binary. The unusually hard
spectrum still needs to be explained.

\cite{2014A&A...570L...2B} also reported the discovery of the QPO in
IGR J173611--4441. They argue that the characteristics of the QPO are
compatible with the tidel disruption event scenario proposed by
\cite{2014MNRAS.444...93D}. \cite{2014A&A...570L...2B} rejected a
neutron-star X-ray binary possibility based on the fact that such
systems hardly show QPOs as such low frequencies. However, we disagree
with this statement since such low frequency QPOs have indeed been
reported for neutron-star systems \citep[see,
e.g.,][]{2007ApJ...660..595L}. Therefore, although we cannot exclude
the tidel disruption event hypothesis, we do consider it more likely
that this system is an unusual accreting neutron-star or black-hole
binary.

\label{lastpage}
\end{document}